\documentclass[german,english]{article}

\pdfoutput=1
\usepackage[latin9]{inputenc}
\usepackage{geometry}
\usepackage{color}
\usepackage{graphicx}
\usepackage{amssymb}
\usepackage{esint}

\newcommand{\lyxaddress}[1]{
\par {\raggedright #1
\vspace{1.4em}
\noindent\par}
}

\usepackage{babel}

\begin{document}

\title{Dephasing by electron-electron interactions in a ballistic Mach-Zehnder
interferometer}

\author{Clemens Neuenhahn and Florian Marquardt}

\maketitle

\lyxaddress{Department of Physics, Arnold Sommerfeld Center for Theoretical Physics,
and Center for NanoScience, Ludwig Maximilians Universität München,
Theresienstr. 37, 80333 Munich, Germany}
\begin{abstract}
We consider a ballistic Mach-Zehnder interferometer for electrons
propagating chirally in one dimension (such as in an integer Quantum
Hall effect edge channel). In such a system, dephasing occurs when
the finite range of the interaction potential is taken into account.
Using the tools of bosonization, we discuss the decay of coherence
as a function of propagation distance and energy. We supplement the
exact solution by a semiclassical approach that is physically transparent
and is exact at high energies. In particular, we study in more detail
the recently predicted universal power-law decay of the coherence
at high energies, where the exponent does not depend on the interaction
strength. In addition, we compare against Keldysh perturbation theory,
which works well for small interaction strength at short propagation
distances.
\end{abstract}
\newcommand{\prb}{Phys. Rev. B}
\newcommand{\prl}{Phys. Rev. Lett.}

\tableofcontents{}

\section{Introduction}

The loss of quantum mechanical phase coherence by a fluctuating environment
plays an essential role in many different branches of modern physics
\cite{2000_Weiss_QuantumDissipativeSystems}. It governs the transition
from the quantum to the classical world \cite{1991_Zurek_PhysicsToday,2003_Joos_Decoherence}
and occurs as an unavoidable consequence of any measurement process
\cite{1992_BraginskyKhalili_QuantumMeasurement}. It introduces the
dephasing time as the time-scale during which it is possible to observe
quantum coherent dynamics in qubits and other quantum objects. Furthermore,
the interference contrast in man-made interferometers is determined
by decoherence. This includes setups for electrons in semiconductors
or metals, for single photons, neutrons, neutral atoms and larger
objects. In each of those examples various different sources of fluctuations
contribute, e.g. thermal radiation, or the interaction with phonons
and other particles.

The paradigmatic setup that is treated in much of the literature on
decoherence is the following: A {}``small'' quantum system with
few degrees of freedom or even a finite-dimensional Hilbert space
(atom, spin, qubit, single particle) interacts with an equilibrium
environment that comprises an infinite number of degrees of freedom,
i.e. a {}``bath''. In the majority of cases, the model of the environment
is restricted even further, to consist of a collection of harmonic
oscillators (e.g. photon or phonon modes). Quantum dissipative systems
of that kind already offer a rich phenomenology, including exponential
or power-law decay of the coherence in time, as well as dissipative
phase transitions at strong coupling \cite{1987_Leggett_ReviewSpinBoson,2000_Weiss_QuantumDissipativeSystems}.

However, there are several situations in which one is forced to go
beyond that class of models. We mention three of the most important
examples, where the nature of the environment has to be reconsidered.
(i) The environment may be driven out of equilibrium (as in the interaction
with electrical currents or a photon stream emanating from a laser),
which is particularly important for measurement setups. As a consequence,
extra fluctuations are introduced and there is no simple relation
any more between the dissipative response and the fluctuations. (ii)
The environment may be different from a bath of harmonic oscillators,
such that the resulting fluctuations are not Gaussian. This comprises
examples like spin baths, fluctuators in solids, or potential fluctuations
being produced by discrete particles, such as electrons. The peculiar
features of non-Gaussian environments may be observed at strong coupling,
and they include oscillations of the interference contrast with time
\cite{2007_05_NederMarquardt_NJP_NonGaussian,2008_05_AbelMarquardt_QuantumTelegraphNoise}.
(iii) It may be hard to cleanly separate (conceptually and technically)
the environment from the system. This is the case for interacting
many-particle systems, where one is interested in the effects on a
single particle that is part of the whole, being indistinguishable
from the other particles.

In this paper, we will address the third case, which is important
in a variety of electronic systems, ranging from dephasing in disordered
conductors to electronic interferometers. Here we will focus on interacting
electrons moving inside a ballistic interferometer, although we also
would like to mention our recent related work on the importance of
Pauli blocking for dephasing in disordered electron systems \cite{2005_10_WeaklocDecoherenceOne,2005_10_WeaklocDecoherenceTwo,2006_04_DecoherenceReview}
. More specifically, we will discuss the loss of coherence in an interacting
one-dimensional chiral fermion system. Besides representing an exactly
solvable model system, where one is not restricted to perturbation
theory in the interaction, this also has become relevant for experiments
recently. In a series of experimental studies \cite{2003_Heiblum_MachZehnder,2006_01_Neder_VisibilityOscillations,2006_07_MZ_DephasingNonGaussianNoise_NederMarquardt,2008_02_Strunk_MZ_Coherence_FillingFactor,2008_03_Roche_CoherenceLengthMZ},
first initiated at the Weizmann institute, an electronic Mach-Zehnder
interferometer has been realized, employing edge channels in the integer
quantum Hall effect. The interference contrast as a function of voltage
and temperature has been analyzed, and only a fraction of the features
have been explained by now. 

On the theoretical side, dephasing in such a setup has been discussed
both for dephasing by external fluctuations \cite{2004_Marquardt_MZ_PRL,2004_10_Marquardt_MZQB_PRL,2004_Marquardt_MZ_PRB,2005_Foerster_MZ_FCS,2005_ChungBuettiker_DephasingShotNoise_MZ,2006_04_MZQB_Long,2006_07_MZ_DephasingNonGaussianNoise_NederMarquardt,2007_05_NederMarquardt_NJP_NonGaussian}
(such as phonons, defect fluctuators or Nyquist noise from external
gates, or {}``dephasing terminal'' reservoirs), as well as by the
intrinsic electron-electron interaction \cite{2001_SeeligBuettiker_MZDephasing,2006_04_LawFeldmanGefen_FQHE_MZ,2007_05_NederMarquardt_NJP_NonGaussian,2006_09_Sukhorukov_MZ_CoupledEdges,2007_ChalkerGefen_MZ,2007_11_NederGinossar_ShotNoiseDephasing,2008_01_Shenja_TwoChannelDephasing}. 

Electron-electron interactions in one-dimensional systems are usually
described within the Luttinger liquid framework. Dephasing of electrons
in Luttinger liquids is interesting as an example of a non-perturbative,
strongly correlated model system, and has been studied already in
a number of works \cite{1983_ApelRice_LLdephasing,2002_LeHur_LLdephasing,2005_LeHur_ElectronFractionalization,2005_Mirlin_DephasingLLweakLoc,2006_LeHur_LifetimeLL,2008_04_GutmanGefenMirlin_NonEquilibriumLL}.\textbf{
}In contrast, the situation for (spinless) chiral interacting fermion
systems, such as edge states in the integer quantum Hall effect (QHE),
seems to be straightfoward. Within the commonly discussed {}``g-ology''
framework, one assumes pointlike interactions. In that case, an interacting
chiral model is nothing but a Fermi gas with a renormalized velocity.
No further interaction effects are present, in particular there is
no dephasing. However, for interferometers in nonequilibrium (at finite
bias voltage) or at finite temperatures, one may still encounter interesting
physics, provided the finite range of the interaction potential\textcolor{red}{
}\textcolor{black}{is accounted} for. This was discussed only recently
\cite{2007_ChalkerGefen_MZ}, although there are some nice and thorough
earlier studies\textbf{ }\cite{1993_MedenSchoenhammer_SpectrumLL,1999_Meden_Luttinger_GF_NonUniversal}\textbf{
}of the momentum-resolved density of states in such systems. 

In the present paper, we build on the recent analysis of Chalker,
Gefen, and Veillette \cite{2007_ChalkerGefen_MZ}. These authors
modeled the interacting Mach-Zehnder interferometer as consisting
of two one-dimensional chiral interacting fermion systems, corresponding
to the two channels or arms of the interferometer. The two channels
are tunnel-coupled weakly at two locations, representing the quantum
point contacts (i.e. the beam splitters) of the experimental setup.
By staying within the regime of weak tunnel coupling (low transmission),
one is able to express the current in lowest order perturbation theory
with regard to the tunneling. The result is a formula for the current
(and, subsequently, the interference contrast or {}``visibility'')
that only involves the Green's functions of the interacting channels
in the absence of tunneling. These Green's functions can be obtained
using the tools of bosonization. In order to obtain nontrivial results,
it is necessary to go beyond the commonly employed assumption of pointlike
interactions. Only when treating the full dependence of the interaction
potential on the distance of the electrons, interaction effects beyond
a simple velocity renormalization are observed. The main results of
their study are that at low voltages and temperatures the interference
contrast becomes perfect, while the suppression of contrast at increasing
voltages and temperatures depends on the details of the interaction
potential. 

In our work, we will first review the general expression for the current
that consists of two parts, a flux-independent term and the interference
contribution that will be suppressed by interaction-induced decoherence
(section \ref{sec:The-electronic-Mach-Zehnder}). In contrast to \cite{2007_ChalkerGefen_MZ},
we formulate the answer in terms of the Green's function in energy-position
space. This has the advantage of corresponding directly to the contribution
of an electron at energy $\epsilon$ that travels a distance $x$
inside one of the arms of the interferometer. Next, we review the
model Hamiltonian and the solution by bosonization (section \ref{sec:Solution-by-bosonization}).
We discuss the general features of the Green's function $\left|G^{>}(\epsilon,x)\right|$
that has been obtained by numerical evaluation of the exact bosonization
expressions, and study the influence of the coupling constant. At
low energies, the decay with propagation distance $x$ is weak. It
becomes faster when the energy rises above the characteristic energy
that characterizes the finite range of the interaction. Finally, at
large energies, the decay becomes independent of energy. 

The latter asymptotic regime is then the subject of a more detailed
analysis using a physically transparent semiclassical picture (section
\ref{sub:Semiclassical-model-of}). This picture of a single electron
interacting with the potential fluctuations produced by the other
electrons has been exploited by us in a recent short paper \cite{2008_06_Neuenhahn_UniversalDephasing_Short}
to discuss dephasing in an interacting chiral fermion system. It had
been introduced earlier to deal with dephasing of ballistically propagating
electrons in contact with an arbitrary quantum environment \cite{2004_10_Marquardt_MZQB_PRL,2006_04_MZQB_Long},
and has also been suggested independently in the context of two interacting
Luttinger liquids \cite{2005_LeHur_ElectronFractionalization}. Here,
we provide more details of the calculation and an extended discussion
of the fluctuation spectrum that is seen by the moving electron in
its frame of reference. One of the main results is that at high energies
there is a {}``universal'' power-law decay $\left|G^{>}(\epsilon,x)\right|\propto1/x$
of the electron's coherence, with an exponent independent of interaction
strength. We also analyze the situation at finite temperature, where
one has to discuss the transformation of the lab-frame temperature
into an effective temperature in the co-moving frame of reference. 

Although in principle, for this particular problem, the bosonization
solution is fully sufficient, we conclude our analysis with a section
on perturbation theory (section \ref{sub:Low-energy--}). This is
done in anticipation of going to different setups where an exact solution
is no longer possible. Even though we will only discuss the equilibrium
Green's function of the interacting system, we employ the nonequilibrium
(Keldysh) diagrammatic technique, to provide for a straightforward
extension to situations where this is needed. We calculate the self-energy
up to second order in the interaction. This includes a diagram that
describes the decay by emission of a plasmonic excitation, which is
however partially cancelled at low energies by an exchange-type diagram.
We will show that at short propagation distances and for modest coupling
strengths, the Keldysh result provides a good approximation to the
exact solution for $\left|G^{>}(\epsilon,x)\right|$, even though
its structure in $(\epsilon,k)$-space is qualitatively different
from the bosonization result.

\section{\label{sec:The-electronic-Mach-Zehnder}The electronic Mach-Zehnder
Interferometer}

The electronic Mach-Zehnder interferometer is one of the simplest
model systems where the interplay of quantum mechanical coherence
and many-body effects can be studied, both in theory and experiment. 

\begin{figure}
\includegraphics[width=1\columnwidth]{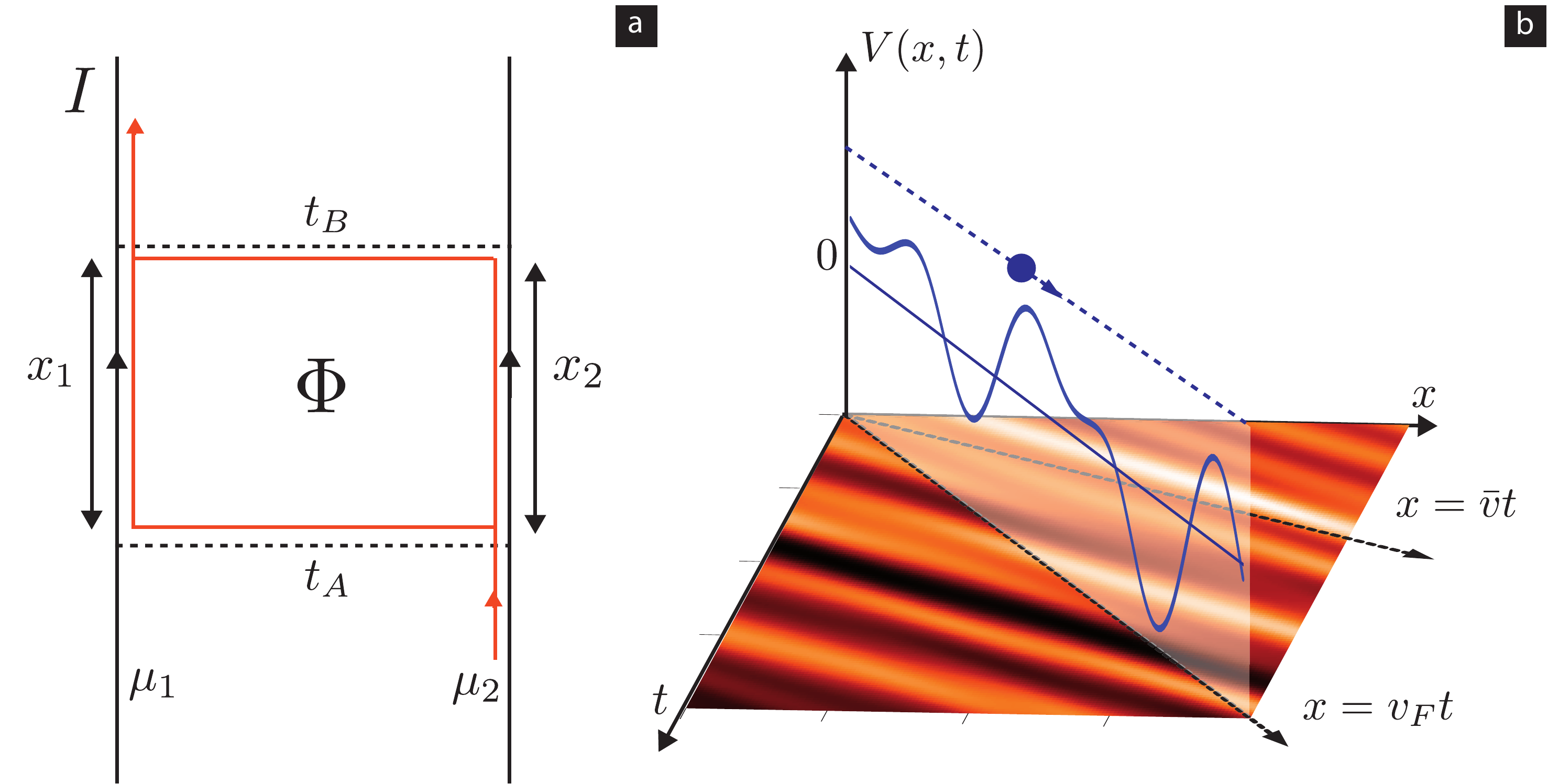}

\caption{(a) Scheme of the interferometer setup. The two channels $1$ and
$2$ of length $x_{1,2}$ and the corresponding chemical potentials
$\mu_{1,2}$ are indicated. The electrons can tunnel at QPCs A and
B, with tunnel amplitudes $t_{A}$ and $t_{B}$. By tuning the magnetic
flux $\Phi$ through the interferometer, one observes an interference
pattern $I(\phi)$. The solid orange lines denote the current through
the interferometer. \protect \\
(b) A single electron propagating at high energies feels a fluctuating
quantum potential $\hat{V}(t)$, due to the interaction with the density
fluctuations in the sea of other electrons. We show a density plot
of the potential, which is produced by the electronic density fluctuations
(plasmons) in the channel. The plasmons are moving with a renormalized
velocity $\bar{v}$ (see main text) while the high-energy electron
moves with the bare Fermi velocity $v_{F}$. It picks up a random
phase, which leads to dephasing. This is the picture underlying the
semiclassical approach in Section \ref{sub:Semiclassical-model-of}.
\label{fig:MZ}}

\end{figure}

To set up the description, we first neglect interactions and imagine
a single electron moving through the interferometer. The interferometer
itself is described as two parallel one-dimensional channels in which
electrons propagate into the same direction (see Fig.\ref{fig:MZ}a).
At two tunnel contacts (i.e. quantum point contacts (QPC's) in the
experimental realization), these channels are coupled by tunneling
amplitudes $t_{A}$ and $t_{B}$. Further below, we will assume these
tunneling probabilities to be small perturbations, coupling lead 1
(left channel) and lead 2 (right channel). Furthermore, a magnetic
flux is enclosed by the interferometer, which leads to an additional
Aharonov-Bohm phase $\phi$. 

In the experiment the current $I$ through the interferometer, i.e.
the current between the two leads, measured at the output port, is
the quantity of interest (see Fig.\ref{fig:MZ}). It contains two
types of contributions: one flux-independent constant term and one
interference term that depends on\textcolor{red}{ }$\cos(\phi)$.
The contrast of the interference fringes observed in $I(\phi)=I_{0}+I_{{\rm {\rm coh}}}(\phi)$
can be quantified via the so-called visibility

\begin{equation}
v_{I}=\frac{I_{{\rm max}}-I_{{\rm min}}}{I_{{\rm max}}+I_{{\rm min}}}\label{eq:}\end{equation}
where $I_{{\rm max}}$ ($I_{{\rm min}}$) are the maximum (minimum)
current as a function of flux. This definition is chosen that the
visibility is equal to one for perfect interference contrast. This
can be used as a direct measure for the coherence of the system. The
coherence can be destroyed by the influence of an external bath as
well as by internal interactions like the Coulomb interaction between
the electrons inside the interferometer. 

Treating the interferometer as a many-body system yields expressions
for the current through the interferometer, which are not as obvious
in a physical sense as in the single particle picture. Therefore the
goal of this section is to formulate the quantities of interest in
the physically most intuitive and transparent way. 

\subsection{\label{sub:Current}Current}

The observable of interest in the present setup is the current through
the interferometer (see Fig.\ref{fig:MZ}a)due to a finite bias voltage
between the two leads, i.e. $\mu_{1}-\mu_{2}\neq0$. Dealing with
the electron-electron interaction exactly using the bosonization technique
has the disadvantage that we have to treat the tunneling between the
channels in perturbation theory. In the following we define the quantum
point contacts A and B to be at the positions $x_{j}^{A}=0$ and $x_{j}^{B}=x_{j}$,
respectively (where $j=1,2$ is the channel index). Then the tunneling
Hamiltonian is given by\begin{equation}
\hat{H}^{T}:=t_{A}\hat{\Psi}_{1}^{\dagger}(0)\hat{\Psi}_{2}^{}(0)+t_{B}\hat{\Psi}_{1}^{\dagger}(x_{1})\hat{\Psi}_{2}^{}(x_{2})+h.c.\label{eq:TunnelingHamiltonian}\end{equation}
The current into channel 1 is ($q_{e}<0$):\begin{equation}
\hat{I}=q_{e}\frac{d}{dt}\hat{N}_{1}.\label{eq:Current}\end{equation}
Thus the Heisenberg equation of motion yields

\begin{equation}
\hat{I}=-q_{e}i[\hat{N}_{1},\hat{H}_{0}+\hat{H}^{T}]=-q_{e}i\left[t_{A}\hat{\Psi}_{1}^{\dagger}(0)\hat{\Psi}_{2}^{}(0)+t_{B}\hat{\Psi}_{1}^{\dagger}(x_{1})\hat{\Psi}_{2}^{}(x_{2})\right]+h.c\quad.\label{eq:current2}\end{equation}
Now we change to the interaction picture with respect to $\hat{H}_{0}$
(where $\hat{H}_{0}$ denotes the interacting electron Hamiltonian
in the absence of tunneling, which we define in section \ref{sec:Solution-by-bosonization}),
setting $\hat{A}_{H_{0}}(t)\equiv e^{i\hat{H_{0}}t}\hat{A}e^{-i\hat{H}_{0}t}$.
We are interested in the steady-state current through the interferometer,
which we obtain as a Kubo-type expression, in linear response with
respect to the tunneling Hamiltonian, at arbitrary bias voltages:\begin{equation}
I=\frac{1}{i}\int_{-\infty}^{0}dt\langle\left[\hat{I}_{H_{0}}(0),\,\hat{H}_{H_{0}}^{T}(t)\right]\rangle.\label{eq:CurrentInLinearResponse}\end{equation}

\subsection{\label{sub:Evaluation-of-the}Evaluation of the current}

Starting from Eq.\ (\ref{eq:CurrentInLinearResponse}) and plugging
in the apropriate definitions an explicit expression for the current
can be found. It will be convenient to express the current in terms
of the unperturbed Green's functions of the chiral electron liquids
in the two channels. These Green's functions will therefore be the
primary object of our subsequent discussions.

\subsubsection{Green's functions}

In particular, we will analyze the particle- and hole-propagators

\begin{eqnarray}
G^{>}(x,t) & \equiv & -i\left\langle \hat{\Psi}(x,t)\hat{\Psi}^{\dagger}(0,0)\right\rangle ,\\
G^{<}(x,t) & \equiv & +i\left\langle \hat{\Psi}^{\dagger}(0,0)\hat{\Psi}(x,t)\right\rangle \end{eqnarray}
where we omit the channel index for brevity, unless needed for clarity.
The Fourier transforms are obtained as $G^{>}(k,\omega)=\int dx\int dt\, e^{-i(kx-\omega t)}G^{>}(x,t)$.
In addition, we will need the spectral density $\mathcal{A}(k,\omega)$,
\begin{equation}
\mathcal{A}(k,\omega)\equiv-\frac{1}{\pi}{\rm Im}[G^{R}(k,\omega)],\end{equation}
where $G^{R}(x,t)=-i\Theta(t)\left\langle \left\{ \hat{\Psi}(x,t),\hat{\Psi}^{\dagger}(0,0)\right\} \right\rangle $.
The energy-dependent tunneling density of states follows as \begin{equation}
\nu(\omega):=\int_{-\infty}^{\infty}dk\,\mathcal{A}(\omega,k)\,.\end{equation}

\subsubsection{Flux-independent part of the current}

The flux-independent part of the current is found using\textcolor{red}{
}Eq.\ (\ref{eq:TunnelingHamiltonian}), (\ref{eq:current2}) and
(\ref{eq:CurrentInLinearResponse}) \begin{equation}
I_{0}=q_{e}(|t_{A}|^{2}+|t_{B}|^{2})\int_{-\infty}^{\infty}dt\,\left[G_{1}^{>}(0,-t)G_{2}^{<}(0,t)-G_{1}^{<}(0,-t)G_{2}^{>}(0,t)\right],\label{eq:I_0Timedomain}\end{equation}
which we rewrite by going to the frequency domain \begin{equation}
I_{0}=q_{e}\left(\left|t_{A}\right|^{2}+\left|t_{B}\right|^{2}\right)\int(d\omega)\left[G_{1}^{>}(0,\omega)G_{2}^{<}(0,\omega)-G_{1}^{<}(0,\omega)G_{2}^{>}(0,\omega)\right],\end{equation}
where $\int(d\omega)\ldots=\int d\omega/2\pi\ldots$. Using the identities\begin{eqnarray}
G_{j}^{>}(x=0,\omega) & = & \int(dk)G_{j}^{>}(k,\omega)=-i\left[1-f_{j}(\omega)\right]\cdot\nu_{j}(\omega)\\
G_{j}^{<}(x=0,\omega) & = & if_{j}(\omega)\cdot\nu_{j}(\omega),\end{eqnarray}
where $f_{j}(\omega)$ denotes the Fermi function. We can reformulate
the expression as a function of the tunneling density of states $\nu(\omega)$, 

\[
I_{0}=q_{e}\left(\left|t_{A}\right|^{2}+\left|t_{B}\right|^{2}\right)\int(d\omega)\nu_{1}(\omega)\nu_{2}(\omega)\left[\underbrace{f_{2}(\omega)[1-f_{1}(\omega)]}_{2\rightarrow1}-\underbrace{f_{1}(\omega)[1-f_{2}(\omega)]}_{1\rightarrow2}\right],\]
which finally yields the most intuitive form describing the sum of
tunneling currents at two point-like locations:\begin{equation}
I_{0}=q_{e}\left(\left|t_{A}\right|^{2}+\left|t_{B}\right|^{2}\right)\int(d\omega)\nu_{1}(\omega)\nu_{2}(\omega)\left[f_{2}(\omega)-f_{1}(\omega)\right].\label{eq:I_0}\end{equation}

In particular, these expressions show that the flux-independent part
of the current only depends on the tunneling density of states. It
is independent of the length of the interferometer arms. This is to
be expected, as that part of the current is insensitive to the electrons'
coherence, and therefore the decay of coherence as a function of propagation
distance will not enter here.

\subsubsection{Coherent part}

The Mach-Zehnder setup is intended to investigate the coherence of
the electron system and therefore the most interesting quantity is
the coherent part of the current, which we define to be the flux-dependent
contribution. Again, using Eq.\ (\ref{eq:TunnelingHamiltonian}),
(\ref{eq:current2}) and (\ref{eq:CurrentInLinearResponse}) it yields
:\begin{equation}
I_{{\rm coh}}(\phi)=q_{e}\int(d\omega)\left[(t_{A}t_{B}^{\ast})e^{-i\phi}\cdot G_{1}^{>}(\omega,x_{1})G_{2}^{<}(\omega,-x_{2})-(t_{A}^{\ast}t_{B})e^{i\phi}\cdot G_{1}^{<}(\omega,-x_{1})G_{2}^{>}(\omega,x_{2})\right]+c.c.\,.\label{eq:I_coh}\end{equation}
At $T=0$, in a situation where the particle current flows from channel
2 to 1, only the first term (and its complex conjugate) contributes. 

It might be helpful to see how the structure of this term $G_{1}^{>}G_{2}^{<}$
can be understood in an intuitive, if slightly imprecise, way, that
also relates to our subsequent semiclassical discussion. When the
full beam in channel 2 impinges onto the first beam-splitter A, we
obtain a superposition between two many-particle states: With an amplitude
near unity, nothing happens (no tunneling takes place), and we denote
this state as $ $$\left|0\right\rangle $. There is a small chance
(of amplitude $t_{A}$) for a particle to tunnel through A into channel
1, leaving behind a hole in channel 2. As time passes, the second
part acquires an amplitude (relative to the first) that is given by
the product of propagation amplitudes for the electron ($\psi_{1}$)
and the hole ($\psi_{2}^{*}$), resulting in:

\begin{equation}
\left|0\right\rangle +t_{A}\psi_{1}\psi_{2}^{*}\left|1_{p},2_{h}\right\rangle \,.\end{equation}
Including the action of the second beam splitter B, and the Aharonov-Bohm
phase, the total probability to detect an extra electron in the output
port (channel 1) is therefore

\begin{equation}
\left|t_{B}e^{i\phi}+t_{A}\psi_{1}\psi_{2}^{*}\right|^{2},\end{equation}
which gives rise to the interference term 

\begin{equation}
t_{A}t_{B}^{*}e^{-i\phi}\psi_{1}\psi_{2}^{*}+c.c..\end{equation}
Averaging the amplitudes over phase fluctuations induced by the interaction,
we arrive at the propagators, replacing $\left\langle \psi_{1}\right\rangle $
by the particle propagator $G_{1}^{>}$, and $\left\langle \psi_{2}^{*}\right\rangle $
by the hole propagator $G_{2}^{<}$. The full analysis keeps track
of energy conservation

Thus, in the many-body picture, the observation of an interference
term in the current is seen to depend both on the passage of an electron
through channel 2 ($G_{2}^{>}$) as well as the coherent propagation
of the corresponding hole, of the same energy $\omega$, in channel
1 ($G_{1}^{<}$). This issue has been discussed before, both for the
Mach-Zehnder interferometer and weak localization \cite{2004_10_Marquardt_MZQB_PRL,2006_04_MZQB_Long,2006_04_DecoherenceReview,2005_10_WeaklocDecoherenceOne}.

\subsection{\label{sub:Visibility}Visibility}

In the Mach-Zehnder setup, the so called visibility is used as a measure
of the coherence of the system. There are different definitions (experimentally,
often the differential visibility is employed). However we will define
the visibility in terms of the total current, as \begin{equation}
v_{I}(V,T)\equiv\frac{{\rm max}_{\phi}I_{coh}(\phi)}{I_{0}}=\frac{I_{{\rm max}}-I_{{\rm min}}}{I_{{\rm max}}^{}+I_{{\rm min}}}\,.\end{equation}
The bias voltage is defined as $\mu_{1}-\mu_{2}=q_{e}V$ and we set
$V>0$. We will also focus on zero temperature, as this seems to be
the most interesting case. Now the visibility can be written in a
compact form (here shown for $T=0$):\begin{equation}
v_{I}=\frac{2|t_{A}t_{B}^{\ast}|}{|t_{A}|^{2}+|t_{B}|^{2}}\cdot\frac{\left|\int_{0}^{|q_{e}V|}d\omega\; G^{>}(\omega,x_{1})\cdot G_{}^{<}(\omega-\left|q_{e}\right|V,-x_{2})\right|}{\int_{0}^{\left|q_{e}V\right|}d\omega\;\nu(\omega)\cdot\nu(\omega-\left|q_{e}\right|V)}\quad(V\geq0).\label{eq:visi_compact}\end{equation}
Note that the channel indices of the Green's functions (GFs) are omitted,
as in this formula the GFs are defined with respect to a fixed density
(and potential $u_{j}=0$) and all the explicit dependence on the
bias voltage is shifted to the GF arguments.

Motivated by the structure of Eq.\ \ref{eq:I_coh} in the following
we will focus the attention on analyzing the function $G_{j}^{>}(\epsilon,x)$
in three different ways. First we will apply the bosonization technique,
i.e. we will include the intrinsic interaction in all orders. The
main disadvantage of the bosonized picture is that we are looking
at single particles tunneling between channels, while we are phrasing
the description in terms of collective, bosonic excitations which
prohibit a simple physical picture of the process of dephasing. Therefore,
as an alternative point of view, we will discuss a transparent semiclassical
model for electrons propagating high above the Fermi energy, subject
to the fluctuations produced by the rest of the electrons. Finally,
we will complement this analysis by studying the system in Keldysh
perturbation theory for the fermions, up to second order in the coupling
strength.

\section{\label{sec:Solution-by-bosonization}Solution by bosonization}

In this section we recall how to obtain the Green's functions needed
here via bosonization \cite{1998_10_DelftSchoeller_BosonizationReview,2006_Giamarchi_Book}.
Experienced readers may skip the section, and can refer to it later
regarding the notation.

\subsection{\label{sub:Hamiltonian}Hamiltonian and formal solution}

We start from interacting right-moving chiral fermions in the two
channels $j=1,2$, employing a linearised dispersion relation (we
set $\hbar=1$):

\begin{equation}
\hat{H}_{0}=\sum_{j=1,2}\left[\sum_{k>0}^{\infty}(u_{j}+v_{F}k)\hat{c}_{j,k}^{\dagger}\hat{c}_{j,k}+\hat{H}_{{\rm int},j}\right]\,.\label{eq:Linearised}\end{equation}
where $v_{F}$ denotes the Fermi velocity and $u_{j}$ is a constant
which fixes the chemical potential of the channel. The particle operators
for the chiral electrons are \begin{equation}
\hat{\Psi}_{j}(x)=\frac{1}{\sqrt{L}}\sum_{k>0}e^{ikx}\hat{c}_{j,k}\,,\end{equation}
where $L$ describes the size of the artificial normalization volume
(with $L\rightarrow\infty$ in the end). We also introduce the density
fluctuations within each channel:\begin{equation}
\hat{\rho}_{j}(x)\equiv\hat{\Psi}_{j}^{\dagger}(x)\hat{\Psi}_{j}(x)-\bar{\rho}_{j}\,\end{equation}
such that the Fourier components are given by

\begin{eqnarray}
\hat{\rho}_{j}(x) & = & \frac{1}{L}\sum_{q\neq0}\hat{\rho}_{q,j}e^{iqx}\\
\hat{\rho}{}_{q,j} & = & \sum_{k>0}\hat{c}_{k,j}^{\dagger}\hat{c}_{k+q,j},\end{eqnarray}
with $\hat{\rho}_{-q,j}=\hat{\rho}_{q,j}^{\dagger}$. The average
density $\bar{\rho}_{j}$ enters $\mu_{j}=u_{j}+2\pi v_{F}\bar{\rho}_{j}$.

As we take the two interferometer channels to be spatially separated,
we only have to take care of intrachannel interactions. Transforming
the interaction Hamiltonian

\begin{equation}
\hat{H}_{{\rm int},j}=\frac{1}{2}\int dx\int dx'\hat{\Psi}_{j}^{\dagger}(x)\hat{\Psi}_{j}^{\dagger}(x')U(x-x')\hat{\Psi}_{j}(x')\hat{\Psi}_{j}(x)\end{equation}
into momentum space yields\begin{equation}
\hat{H}_{{\rm int},j}=\frac{1}{2L}\sum_{k>0,k'>0,q}U_{q}\hat{c}_{k+q,j}^{\dagger}\hat{c}_{k'-q,j}^{\dagger}\hat{c}_{k',j}\hat{c}{}_{k,j},\end{equation}
where $U_{q}=\int dx\, e^{-iqx}U(x)$ are the Fourier components of
the interaction potential.

We construct bosonic operators from the Fourier components of the
density in the standard way. As we are only dealing with chiral electrons,
we only need to consider $q>0$ in the following:\begin{eqnarray}
(q>0)\quad\hat{b}_{q,j}=\left(\frac{2\pi}{Lq}\right)^{1/2}\hat{\rho}_{q,j} &  & \hat{b}_{q,j}^{\dagger}=\left(\frac{2\pi}{Lq}\right)^{1/2}\hat{\rho}{}_{-q,j}\,.\label{eq:}\end{eqnarray}
These operators fulfill\begin{equation}
[\hat{b}_{j,q},\hat{b}_{j',q'}^{\dagger}]=\delta_{j,j'}\delta_{q,q'}\,.\end{equation}
As usual, the main advantage of bosonization consists in being able
to write the kinetic part of the Hamiltonian as a quadratic form in
boson operators:\begin{equation}
\hat{H}_{0}=\sum_{j=1,2}\left[v_{F}\sum_{q>0}q\hat{b}_{q,j}^{\dagger}\hat{b}_{q,j}+\mu_{j}\hat{N}_{j}\right]+{\rm const}\,.\label{eq:-1}\end{equation}
The interaction part of the Hamiltonian reads:\[
\]
\begin{equation}
\hat{H}_{{\rm int},j}=\frac{1}{L}\sum_{q>0}U_{q}\hat{\rho}_{q,j}^{\dagger}\hat{\rho}{}_{q,j}\,.\end{equation}
Thus, the Hamiltonian is already in diagonal form,\begin{equation}
\hat{H}=\sum_{j=1,2}\left[\sum_{q>0}\omega(q)\hat{b}_{q,j}^{\dagger}\hat{b}_{q,j}+\mu_{j}\hat{N}_{j}\right],\end{equation}
with the plasmonic dispersion relation of an interacting chiral 1D
electron system:\begin{equation}
\omega(q)=v_{F}q\left[1+\frac{U_{q}}{2\pi v_{F}}\right]\label{eq:DispersionRelationBosoni}\end{equation}
For the following discussions, we introduce the dimensionless coupling
constant $\alpha=\frac{U(q\rightarrow0)}{2\pi v_{F}}$, where $\alpha\in]-1,\infty[$.
The renormalized plasmon velocity at small wavenumbers is $\bar{v}=v_{F}(1+\alpha)$.
Negative values of the coupling constant are related to attractive
interations, positive values to repulsion (at small wavenumbers).
For $\alpha\rightarrow-1$ the plasmon velocity tends to zero, $\bar{v}\rightarrow0$.
For $\alpha<-1$ the system is unstable, i.e. formally $\omega(q)<0$
for $q>0$.

The final step is to express the single-particle operators using bosonic
fields \cite{2006_Giamarchi_Book,1998_10_DelftSchoeller_BosonizationReview}
(assuming $L\rightarrow\infty)$:\begin{equation}
\hat{\Psi}_{j}(x)=\frac{\hat{F}_{j}}{\sqrt{2\pi a}}e^{ik_{F}x}e^{-i\hat{\Phi}_{j}(x)}\quad\hat{\Phi}_{j}(x)=i\sum_{q>0}\sqrt{\frac{2\pi}{Lq}}e^{-aq}\left[\hat{b}_{q,j}e^{iqx}-h.c.\right]\,.\label{eq:BosonizedSPO}\end{equation}
The ultraviolet cutoff length $a$ is sent to zero at the end of the
calculation. The Klein operator $\hat{F}_{j}$ annihilates a fermion
in a spatially homogeneous way, with the following commutation relations:\begin{equation}
\{F_{i},F_{j}^{\dagger}\}=\delta_{i,j};\,\{F_{i},F_{j}\}=\{F_{i}^{\dagger},F_{j}^{\dagger}\}=0\,.\end{equation}
The explicit time dependence of the Klein operators is obtained by
using the Heisenberg equation of motion, i.e. $d\hat{F}_{j}/dt=-i[\hat{F}_{j},\hat{H}]=-i\mu_{j}\hat{F}_{j}\Rightarrow\hat{F}_{j}(t)=e^{-i\mu_{j}t}\hat{F}{}_{j}(0)$.
In the end taking into account the relation $k_{F,j}=\frac{2\pi}{L}\bar{N}_{j}$
we arrive at\begin{equation}
\hat{\Psi}_{j}(x)=\hat{F}_{j}\frac{1}{\sqrt{2\pi a}}e^{-i\mu_{j}(t-x/v_{F})}\cdot e^{-i\hat{\Phi}_{j}(x,t)}\,.\label{eq:BosonizedSPO2}\end{equation}

\subsection{Green's function from bosonization}

Now we are able to evaluate the Green's function defined above explicitly
using the bosonized single particle operators $\hat{\Psi}_{}$. The
calculation is done quickly using the fact, that the Hamiltonian in
terms of the bosonic operators is quadratic, i.e. the field $\hat{\Phi}[\hat{b},\hat{b}^{\dagger}]$
can be treated like a Gaussian (quantum) variable, resulting in

\begin{equation}
G^{>}(x,t)=\frac{-i}{2\pi a}e^{-i\mu(t-x/v_{F})}\exp[\left\langle \hat{\Phi}(x,t)\hat{\Phi}(0,0)\right\rangle -\left\langle \hat{\Phi}(0,0)^{2}\right\rangle ],\end{equation}
and analogously for $G^{<}$. By factoring off the non-interacting
Green's function $G_{0}^{>/<}$, we can write 

\begin{equation}
G^{>/<}(x,t)=e^{-i\mu_{}[t-x/v_{F}]}\cdot G_{0}^{>/<}(x,t)\cdot\exp[S_{R}(x,t)\mp iS_{I}(x,t)]\,,\end{equation}
where\begin{equation}
G_{0}^{>/<}(x,t)=\frac{1}{2\beta v_{F}}\cdot\frac{1}{\sinh[\frac{\pi}{\beta v_{f}}(x-v_{F}t\pm i0^{+})]}\,\end{equation}
with $\beta\equiv\frac{1}{T}$ and $k_{B}\equiv1$. All the effects
of the interaction now are included in the exponent where we have
to subtract the non-interacting contribution: \begin{eqnarray}
S_{R} & = & \int_{0}^{\infty}\frac{dq}{q}\{\underbrace{\coth[\frac{\beta\omega_{q}}{2}]\left[\cos[\omega_{q}t-qx]-1\right]}_{\tilde{S_{R}}(\omega_{q})}-\tilde{S}_{R}(\omega_{q}\rightarrow qv_{F})\}\label{eq:SR}\\
S_{I} & = & \int_{0}^{\infty}\frac{dq}{q}\{\underbrace{\sin[\omega_{q}t-qx]}_{\tilde{S}_{I}(\omega_{q})}-\tilde{S}_{I}(\omega_{q}\rightarrow qv_{F})\}.\label{eq:SI}\end{eqnarray}

\subsection{Discussion: Green's function in space and time\label{sub:Discussion:-Green's-function}}

\begin{figure}[p]
\includegraphics[width=1\columnwidth]{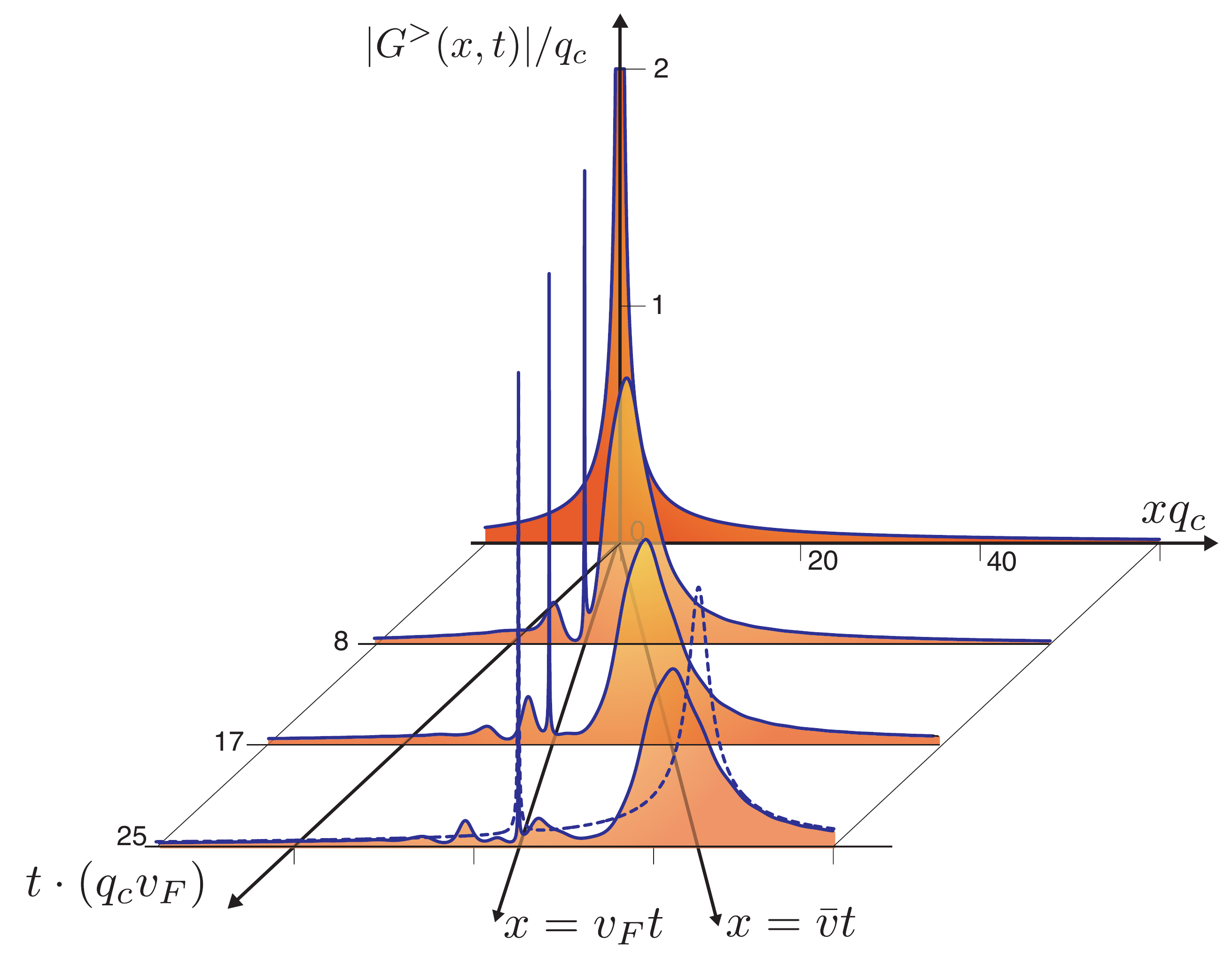}\caption{Numerical evaluation of $\left|G^{>}(x,t)\right|$ at zero temperature
$T=0$ resulting from bosonization, as a function of $x$ and $t$
(solid blue lines). The weight of the sharp peak at $x=v_{F}t$ decreases
for increasing propagation times $t$. The dashed blue line shows
an approximation of the Green's function (see Eq.\ (\ref{eq:Approx}))
which yields good qualitative agreement with the full solution. The
plot is done for $U_{q}=U_{0}e^{-(q/q_{c})^{2}}$with $\frac{U_{0}}{v_{F}}_{}=2\pi\alpha=5$
\label{GXT}}

\end{figure}

Here we discuss the Green's function as a function of space and time.
A more detailed discussion can be found in \cite{2007_ChalkerGefen_MZ}.
The absolute value of the Green's function $G^{>}(x,t)$ is shown
in Fig.\ref{GXT}, at zero temperature (to which we restrict our discussion).\textcolor{black}{
In the following all the numerical evaluations are performed using
a generic smooth interaction potential $U_{q}^{}=U_{0}e^{-(|q|/q_{c})^{s}}$.
We note that all the qualitative results are valid for potentials
which are finite at zero momentum ($U_{0}\neq0$) and which are cut
off beyond some momentum scale. }Those assumptions are not restrictive
and for example are fulfilled for a Coulomb potential with screening
in a quasi one-dimensional channel of finite width. 

In Fig.\ref{GXT}, we observe as the main feature that the Green's
function splits into two parts during its propagation. One of those
propagates with the bare Fermi velocity $v_{F}$ and represents the
unperturbed Green's function, i.e. the high energy part. For increasing
time its weight decreases, i.e. the amplitude of the bare electron
to arrive at $x$ without being scattered decreases. The other peak
represents the low energy part, stemming from energies below $\epsilon-\mu\sim v_{F}q_{c}$.
It moves with the renormalized velocity $\bar{v}$.

We can obtain this structure of $G^{>}(x,t)$ from a crude approximation.
Namely for $T=0$ in a first approximation we can cut the momentum
integral due to the fact that for $q\gg q_{c}$ the integrand vanishes,
i.e.\[
S[x,t]\equiv S_{R}-iS_{{\rm I}}\approx\int_{0}^{\infty}\frac{dq}{q}e^{-q/q_{c}}\left[\exp[-iq(\bar{v}t-x)]-1\right]-\int_{0}^{\infty}\frac{dq}{q}e^{-q/q_{c}}\left[\exp[-iq(v_{F}t-x)]-1\right].\]
The integrals are known and yield\begin{equation}
S[x,t]\approx\ln\left[\frac{x-v_{F}t+iq_{c}^{-1}}{x-\bar{v}t+iq_{c}^{-1}}\right].\end{equation}
Therefore the structure of the Green's function is given by\begin{equation}
G_{T=0}^{>}(x,t)\approx\frac{1}{x-v_{F}t+i0^{+}}\cdot\left[\frac{x-v_{F}t+iq_{c}^{-1}}{x-\bar{v}t+iq_{c}^{-1}}\right],\label{eq:Approx}\end{equation}
displaying both the $\delta$ peak at $x=v_{F}t$ and the broadened
peak at $x=\bar{v}t$, whose width is set by $q_{c}^{-1}$ . In Fig.\ref{GXT}
one can observe the fairly good agreement between the full result
and this first approximation.

\subsection{Green's function vs. position and energy}

\begin{figure}[!t]
\includegraphics[width=1\columnwidth]{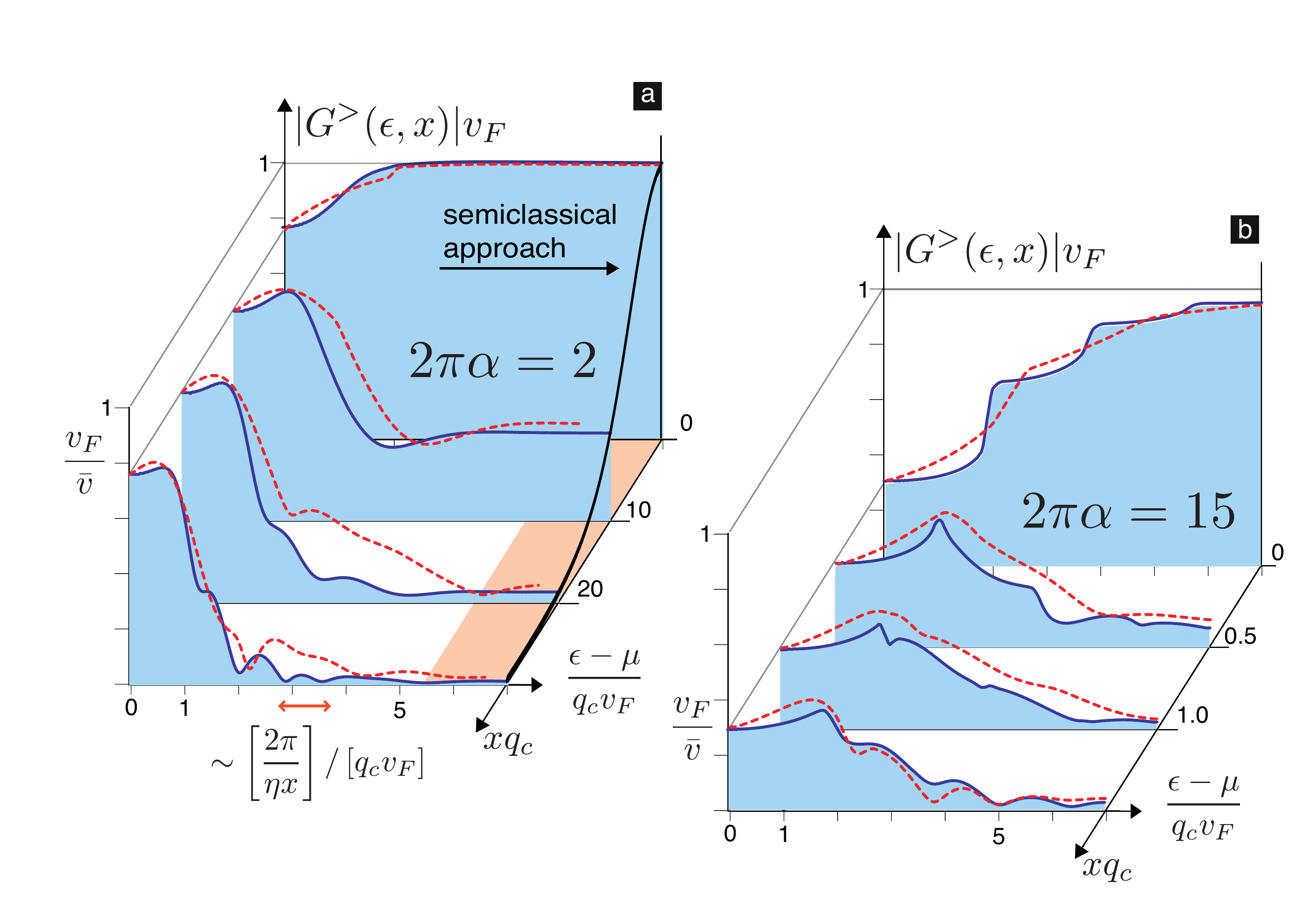}

\caption{a) The electron Green's function $|G^{>}(x,\epsilon)|$, for different
lengths $x$ as a function of energy $\epsilon$, evaluated using
bosonization, for $U(q)=U_{0}e^{-(q/q_{c})^{2}}$ {[}solid blue lines{]}
and for $U(q)=U_{0}e^{-|q/q_{c}|}$ {[}dashed red lines{]}. At high
energies we show the result coming from the semiclassical approach
for the Gaussian potential {[}solid black line{]}. The red area indicates
the regime of validity for the semiclassical (high-energy) approximation.$ $
The interaction strengths are: $2\pi\alpha=2$ (a) and $2\pi\alpha=15$
(b). In the high-energy limit, the semiclassical solution is valid
for arbitrary coupling strength. However, the energies for which the
description is valid become larger for increasing coupling strength.
In $ $(b) this limit is beyond the presented energy interval. Therefore
here we do not show the semiclassical solution. \label{gexfig}}

\end{figure}

As shown above in Eq.~(\ref{eq:I_coh}) and (\ref{eq:visi_compact}),
the current through the interferometer is determined by the propagators
$G^{>/<}(\epsilon,x)$. Therefore, in the following our main focus
will be on this function, which can be thought of as the amplitude
for an electron of energy $\epsilon$ to propagate unperturbed over
a distance $x$.

The function is shown in Fig.~\ref{gexfig}, where we plot the numerical
evaluation of the exact result obtained using the bosonization technique.
This is done for two values of the coupling strength $\alpha$ and
for different interaction potentials. There are some main features
which can be observed by having a brief look at Fig.~\ref{gexfig},
while for a detailed discussion we refer to the following section.

(i) At $x=0$, where $\left|G^{>}(x=0,\epsilon)\right|=\nu(\epsilon)$
equals the tunneling density of states, there is a finite dip at low
energies. This is a static interaction effect. For repulsive interactions
it represents the suppression of the tunneling density by a factor
$v_{F}/\bar{v}$, due to the interaction-induced increase of the velocity
$\bar{v}$. At high energies ($\epsilon\gg v_{F}q_{c}$), the non-interacting
density of states is recovered.

(ii) At any fixed energy $\epsilon$, the Green's function decays
with increasing propagation length $x$. The shape of the decay (as
a function of $x$) becomes independent of energy for high energies.
In contrast, the decay is suppressed for energies below $\epsilon\sim v_{F}q_{c}$,
and there is no decay in the limit $\epsilon\rightarrow0$. The decay
of the GF is equivalent to dephasing (since in our model there are
no interbranch interactions and correspondingly no vertex corrections).
As a consequence, the absence of decay at zero energy will lead to
perfect visibility at $T=0,V\rightarrow0$. 

(iii) At larger $x$, there are oscillations in the Green's function.
These result from the double-peak structure in the time-domain, with
peaks at $x=vt$ and $x=\bar{v}t$. These lead to a beating term $\exp[i\omega x(v^{-1}-\bar{v}^{-1})]$
in $\left|G^{>}(x,\omega)\right|$. Therefore, the period of oscillations
in the energy domain is determined by the difference between the bare
and the renormalized velocity (see Fig.\ref{gexfig}), viz.: \begin{equation}
\delta\epsilon\approx\frac{2\pi}{x\eta};\,{\rm with}\,\eta=\frac{1}{v}-\frac{1}{\bar{v}}.\end{equation}

\subsection{Large coupling constants}

In this section we want to discuss briefly the shape of the Green's
function in terms of the coupling strength. We emphasize that, once
the shape of the interaction potential is given, the only dimensionless
parameter left is the coupling constant $\alpha=U_{0}/(2\pi v_{F})$.
All the other parameters may be absorbed into a rescaling of the result. 

\begin{figure}
\includegraphics[width=1\columnwidth]{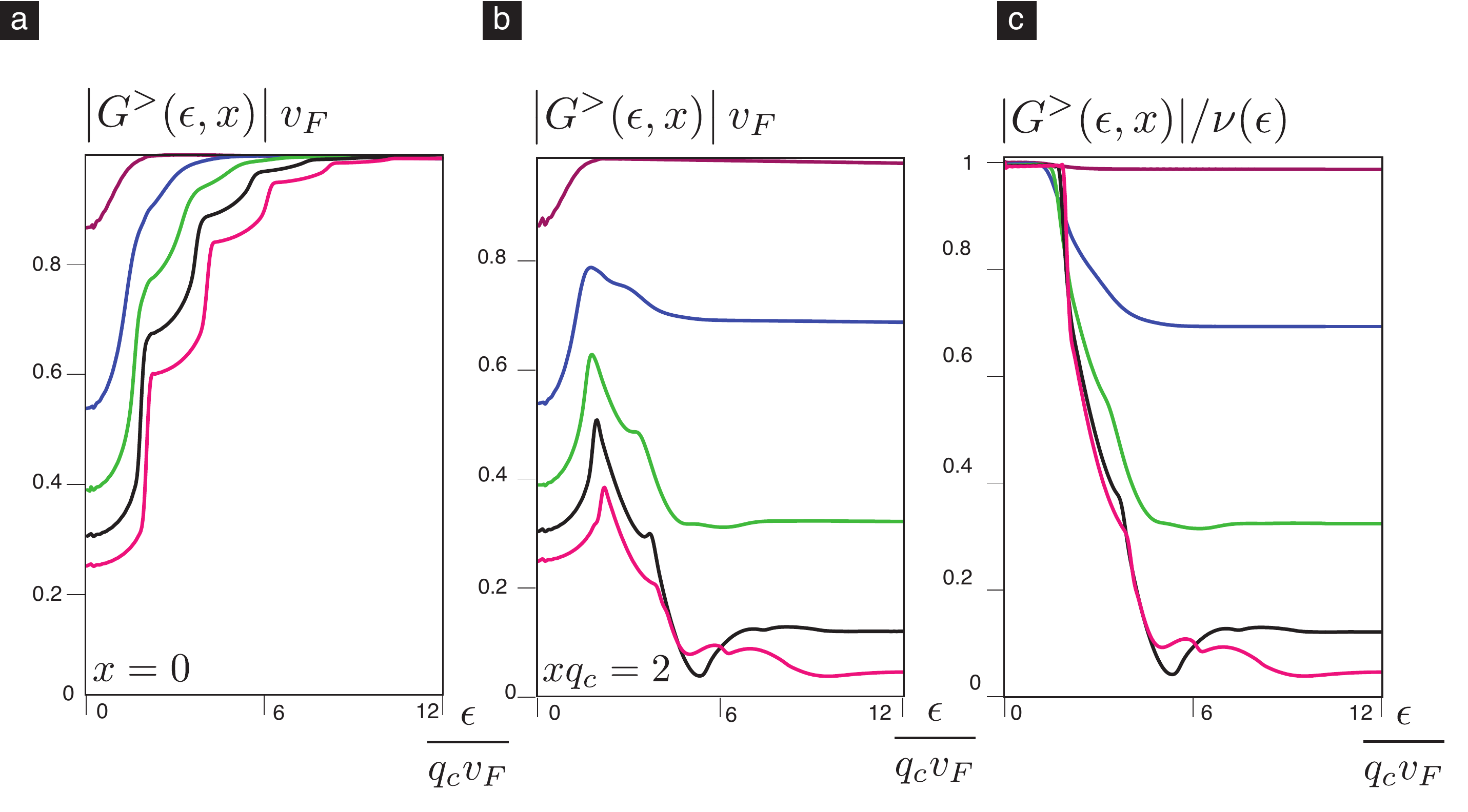}

\caption{Green's function at zero temperature $T=0$ as a function of energy
$\epsilon$ for various coupling strengths $\alpha$, at $x=0$ (a)
and $xq_{c}=2$ (b). \protect \\
c) $|G^{>}(\epsilon,x)|$ for $xq_{c}=2$, divided by the tunnel
density of states $\nu(\epsilon)=|G^{>}(\epsilon,x=0)|$. This might
be interpreted as the electron's coherence as a function of propagation
energy and distance.\protect \\
Here the potential is $U_{q}=2\pi\alpha v_{F}e^{-(q/q_{c})^{2}}$
, where the various values of $\alpha$ are (from top to bottom):
$\alpha=0.16,0.9,1.6,2.3,3$ \label{fig:Large-Coupling-constants}. }

\end{figure}

In Fig. \ref{fig:Large-Coupling-constants} we show $|G^{>}(\epsilon,x)|$
as a function of energy for various coupling strengths (different
curves), both at $x=0$ and at some finite propagation distance $x\neq0$.
For small coupling $\alpha>0$, we just observe the suppression of
the tunneling density of states discussed above. Upon increasing the
coupling strength, a series of rounded steps emerges, suppressing
the tunneling density even further. The same features can be seen
in the shape of the GF at finite $x$, though there they are superimposed
by the decay (describing decoherence) and the oscillations as a function
of energy (discussed in the preceding section). To identifiy the oscillations
in energy which we observe even for small coupling strength in Fig.\ref{fig:Large-Coupling-constants}c
we divide the GF for $x\neq0$ showed in Fig.\ref{fig:Large-Coupling-constants}b
by the tunnel density of states. As expected, those oscillations are
robust against a change in the coupling strength. We have not found
any simple analytical model to discuss the structures observed here.
However, note that in Fig.\ref{gexfig}b one observes that the step
structure is more pronounced for the Gaussian potential compared to
the results for the exponential shape. That shows the strong influence
of the shape of the interaction potential on the step structure. 

We note that the previous discussion in the $(\epsilon,k)$-space
(as opposed to $(\epsilon,x)$) had found non-analytic structures
for the case of a box-shape potential $U_{q}$ \cite{1993_MedenSchoenhammer_SpectrumLL}.

\section{\label{sub:Semiclassical-model-of}Semiclassical model of dephasing }

Up to now we have applied the bosonization technique in order to get
exact information about the decoherence the electron suffers while
passing through the interferometer. However we actually do not know
what is going on in more physical terms. First of all, we do not know
in detail how to distinguish between the single electron we are considering
while traveling through the interferometer and the bath electrons
which are present at the same time. In a Fermi liquid there is no
question about the nature of the single electron, i.e. in the vicinity
of the Fermi edge it can be described as a quasiparticle. In contrast,
in the bosonization approach everything is described in terms of collective
bosonic excitations, i.e. density fluctuations. Therefore, as soon
as the electron tunnels into the interacting system, there is no way
to trace this special electron any longer. This fact leads to difficulties
in understanding the decoherence intuitively, since in the moment
of tunneling the coherent phase information is encoded into the bosonic
degrees of freedom. Fortunately, it turns out that indeed it is possible
to find simple physical pictures which are helpful in understanding
the process of dephasing in more detail besides the mathematical solution. 

For energies much higher than the Fermi energy (to be made more precise
below), it turns out that it is possible to describe the interaction
of a single propagating electron with all the other electrons, by
viewing them as a bosonic quantum bath. To see this, in section \ref{sub:Semiclassical-model-of}
we apply an intuitive, semiclassical method which is able to reproduce
exactly the results from the full bosonization formalism in the high
energy limit.

In a recent work \cite{2008_06_Neuenhahn_UniversalDephasing_Short}
we have already briefly reported on universal dephasing for high-energy
electrons at long distances for $T=0$, based on the semiclassical
approach to be discussed in more detail in the following.

\subsection{Semiclassical approach to the Green's function}

Electrons at high energies $\epsilon$ propagate at the unperturbed
speed $v_{F}$, as can be observed from the corresponding limiting
behaviour of the plasmonic dispersion relation. The decoherence, i.e.
the Green's function, can be described in a transparent semiclassical
framework, that becomes exact in the limit of high energies. We will
confirm this later by comparing against the full bosonization solution.
Within this semiclassical picture, one thinks of the electron propagating
through the channel, while accumulating an additional phase due to
the interaction with the bath formed by all the other electrons. To
model the effective, bosonic bath acting on the single electron, we
make use of the plasmonic dispersion relation which was derived using
the full bosonization technique (Eq.\ \ref{eq:DispersionRelationBosoni}).
As the electron we consider is flying high above the Fermi sea, we
can neglect the backaction of the electron onto the bath. In this
picture the electron only experiences the intrinsic fluctuations of
the bath. The potential acting on such a single high-energy electron
is obtained by convoluting the density fluctuations with the interaction
potential:

\begin{equation}
\hat{V}(t)=\int dx'\, U(x'-v_{F}t)\hat{\rho}(x',t).\label{eq:VUrho}\end{equation}
Note that this definition implies, that the effective potential fluctuations
$\hat{V}(t)$ experienced by the single electron are just the fluctuations
of the bath evaluated at the classical electron position $x=v_{F}t$
at time $t$. This is why we call this model {}``semiclassical''.

If we were dealing with a classical fluctuating potential $V(t)$,
the electron would simply pick up a random phase $\varphi(t)=-\int_{0}^{t}dt'\, V(t')$.
In that case the non-interacting Green's function would have to be
multiplied by a factor $\left\langle e^{i\varphi(t)}\right\rangle $
to obtain the correct Green's function. However, if the quantum nature
of the bath becomes important one has take care of the non-commutativity
of the operator $\hat{V}(t)$ at different times. This can be done
by introducing a time-ordering symbol:

\begin{eqnarray}
e^{-F(t)} & \equiv & \left\langle \hat{T}\exp\left[-i\int_{0}^{t}dt'\hat{V}(t')\right]\right\rangle \nonumber \\
 & = & \exp\left[-\frac{1}{2}\int_{0}^{t}dt_{1}\int_{0}^{t}dt_{2}\,\left\langle \hat{T}\hat{V}(t_{1})\hat{V}(t_{2})\right\rangle \right].\label{eq:expF}\end{eqnarray}
The time $t=x/v_{F}$ in Eq.~(\ref{eq:expF}) is determined by the
propagation length. This is actually identical to the decay of coherence
of a single level whose energy fluctuates. In various contexts, this
is known as the {}``independent boson model'' \cite{2000_Mahan,2002_Marquardt_AB_PRB},
or the case of {}``pure dephasing'' in a (longitudinally coupled)
spin-boson model \cite{2000_Weiss_QuantumDissipativeSystems}. 

We note that the same kind of approach to dephasing of ballistically
propagating electrons has been introduced previously, both for a situation
with a general quantum bath \cite{2004_10_Marquardt_MZQB_PRL,2006_04_MZQB_Long,2006_04_DecoherenceReview},
as well as for two coupled Luttinger liquids \cite{2005_LeHur_ElectronFractionalization}.

Furthermore we note that the decay is independent of energy $\epsilon$.
This is because the propagation speed is energy-independent, and the
distance to the Fermi edge becomes unimportant at high energies as
well. Qualitatively, we have seen this feature before in our discussion
of the full bosonization solution.

In summary, the decay of coherence, described by $F(t)$, is completely
determined by the fluctuation spectrum $\left\langle \hat{V}\hat{V}\right\rangle _{\omega}=\int dt\, e^{i\omega t}\left\langle \hat{V}(t)\hat{V}(0)\right\rangle $
of the potential seen by the electron in the moving frame. To proceed
further we express the time-ordered correlator $\left\langle \hat{T}\hat{V}(t_{1})\hat{V}(t_{2})\right\rangle $
as a sum of commutator and anti-commutator part:$ $\begin{equation}
\left\langle \hat{T}\hat{V}(t_{1})\hat{V}(t_{2})\right\rangle =\frac{1}{2}\left[\left\langle \{\hat{V}(t_{1}),\hat{V}(t_{2})\}\right\rangle +{\rm sgn}(t_{1}-t_{2})\left\langle [\hat{V}(t_{1}),\hat{V}(t_{2})]\right\rangle \right].\label{eq:}\end{equation}
The real part of $F(t)$ and therefore the decay of the Green's function
depends on the symmetrised part of the correlator. This part is formally
similar to the correlator of classical noise, though it also contains
the zero-point fluctuations of the plasmon field:\begin{equation}
{\rm Re}[F(t)]=\frac{1}{4}\int_{0}^{t}dt_{1}\int_{0}^{t}dt_{2}\,\left\langle \{\hat{V}(t_{1}),\hat{V}(t_{2})\}\right\rangle =\int_{-\infty}^{+\infty}\frac{d\omega}{2\pi}\frac{\sin^{2}(\omega x/2v_{F})}{\omega^{2}}\left\langle \{\hat{V},\hat{V}\}\right\rangle _{\omega}.\label{eq:ReF}\end{equation}
\textcolor{black}{In addition, a phase $-{\rm Im}F(t)$ shows up in
the exponent. It is due to the commutator of $\hat{V}$, and thus
it represents a purely quantum mechanical contribution. In terms of
the Fourier transform of the spectrum, this yields}

\begin{eqnarray}
 & {\rm Im}[F(t)]=-\frac{i}{4}\int_{0}^{t}dt_{1}\int_{0}^{t}dt_{2}\, sgn(t_{1}-t_{2})\left\langle [\hat{V}(t_{1}),\hat{V}(t_{2})]\right\rangle \nonumber \\
 & =-\frac{1}{2}\int_{-\infty}^{+\infty}\frac{d\omega}{2\pi}\,\left[\frac{t}{\omega}-\frac{\sin(\omega t)}{\omega^{2}}\right]\left\langle [\hat{V},\hat{V}]\right\rangle _{\omega} & .\label{eq:ImF}\end{eqnarray}

\begin{figure}
\includegraphics[width=1\columnwidth]{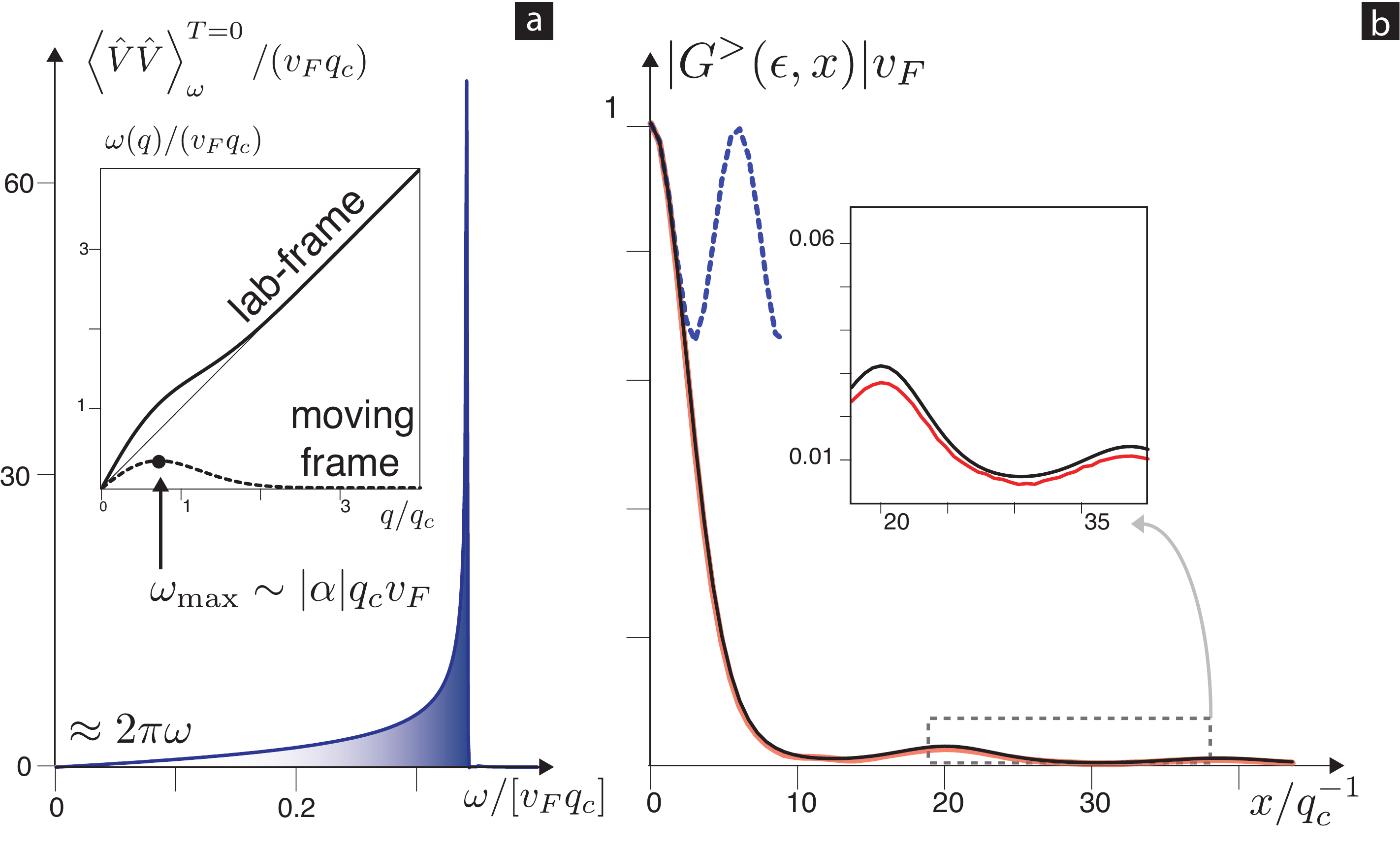}

\caption{a) Plot of the effective spectrum $\langle VV\rangle_{\omega}^{T=0}$
for $\alpha>0$ of the plasmonic bath at the particle position which
moves with velocity $v_{F}$. The spectrum is linear in $\omega$
for small frequencies and diverges like $\frac{1}{\sqrt{\omega_{{\rm max}}-\omega}}$
when approaching $\omega_{{\rm {\rm max}}}$ . The inset shows the
dispersion relation of the plasmonic bath in the laboratory frame
of reference as well as in the electron frame of reference, where
it is getting tilted.{[}dashed curve{]}\protect \\
b) $G^{>}(\epsilon,x)$ as a function of $x$ for large energies
$\epsilon\gg q_{c}v_{F}$ . The solid orange line denotes the numerical
evaluation of the bosonization result, while the solid black line
shows the semiclassical result. For a better comparison of the results,
the inset shows a blow-up of the oscillations. The small deviations
are due to finite numerical precision. In addition, the dashed line
denotes the solution of the Keldysh pertubation theory. All the plots
are done for $U_{q}=U_{0}e^{-(q/q_{c})^{2}}$ with $\frac{U_{0}}{v_{F}}=2\pi\alpha=5$.\label{dispersionrelation}}

\end{figure}

From Eq.~(\ref{eq:VUrho}), we obtain for the potential spectrum
in the co-moving frame\textcolor{red}{ }

\begin{equation}
\left\langle \hat{V}\hat{V}\right\rangle _{\omega}=\int\frac{dq}{2\pi}\,\left|U_{q}\right|^{2}\left\langle \hat{\rho}\hat{\rho}\right\rangle _{q,\omega+v_{F}q}\,.\label{eq:VVomega}\end{equation}
The argument $\omega+v_{F}q$ indicates that we are dealing with the
Galileo-transformed spectrum of the density fluctuations. As a result,
the spectrum of the density fluctuations gets tilted compared to the
original dispersion relation (see Fig.~\ref{dispersionrelation}).
Making use of the plasmonic dispersion $\omega(q)$, which we obtained
from the bosonization method, the density-density correlator yields\begin{equation}
\left\langle \hat{\rho}\hat{\rho}\right\rangle _{q,\omega}=\frac{|q|}{1-e^{-\beta\omega}}\cdot\left[\Theta(q)\delta(\omega-\omega(|q|))-\Theta(-q)\delta(\omega+\omega(|q|))\right].\label{eq:DichteKorrelator}\end{equation}
We obtain

\begin{eqnarray}
 & \left\langle [\hat{V},\hat{V}]\right\rangle _{\omega}=\int\frac{dq}{2\pi}\,|U_{q}|^{2}\left\langle [\hat{\rho},\hat{\rho}]\right\rangle _{q,\omega+qv_{F}}\label{eq:VV_commutator}\\
 & \left\langle \{\hat{V},\hat{V}\}\right\rangle _{\omega}=\int\frac{dq}{2\pi}\,|U_{q}|^{2}\coth\left(\frac{\beta(\omega+qv_{F})}{2}\right)\left\langle [\hat{\rho},\hat{\rho}]\right\rangle _{q,\omega+v_{F}q} & .\label{eq:VV_sym}\end{eqnarray}
The symmetrized correlator can be written as \begin{equation}
\omega>0:\quad\left\langle \{\hat{V},\hat{V}\}\right\rangle _{\omega}=\int_{0}^{\infty}\frac{dq}{2\pi}\,|U_{q}|^{2}q\coth\left(\frac{\beta\omega(q)}{2}\right)\delta(\omega-\frac{|U_{q}q|}{2\pi}).\label{eq:VV_Sym_inv_sign}\end{equation}
 Plugging in these correlators in Eq.\ (\ref{eq:ReF}) and Eq.\ (\ref{eq:ImF}),
finally yields $F(t)$: \begin{equation}
{\rm Re}[F(t)]=\int_{0}^{\infty}\frac{dq}{q}\,\coth\left(\frac{\beta\omega(q)}{2}\right)\left[1-\cos\left(\frac{U_{q}q}{2\pi}t\right)\right]\label{eq:ReF_final}\end{equation}
\begin{equation}
{\rm Im}[F(t)]=\int_{0}^{\infty}\frac{dq}{q}\,\sin\left(\frac{U_{q}q}{2\pi}t\right)-\int_{0}^{\infty}dq\,\frac{U_{q}}{2\pi}\cdot t\,.\label{eq:ImF_final}\end{equation}
Indeed, up to an additional energy renormalization $-\int_{0}^{\infty}dq\,\frac{U_{q}}{2\pi}$,
the resulting Green's function $G^{>}(x=v_{F}t,t)=e^{-F(t)}G_{0}^{>}(x=v_{F}t,t)$
is identical to the one stemming from the bosonization technique for
$x=v_{F}t$ {[}compare to Eqs. (\ref{eq:SR}) and (\ref{eq:SI}){]}.
This constant energy shift had been incorporated into a redefinition
of the chemical potential, as noted in the section on bosonization. 

To be precise, Eqs. (\ref{eq:ReF_final},\ref{eq:ImF_final}) show
that the decoherence of electrons moving along the trajectory $x=v_{F}t$
can be described exactly in terms of the semiclassical approach (which
is based on the assumption of high-energetic electrons). In order
to prove the statement that the decoherence at large energies can
be understood within the semiclassical description, one has to confirm
the initial assumption that high energetic electrons actually move
with the bare Fermi velocity $v_{F}$. Already, in section \ref{sub:Discussion:-Green's-function}
we observed that the Green's function $G^{>}(x,t=\frac{x}{v_{F}})$
determines the high-energy regime of the Fourier transform $G^{>}(x,\epsilon)$.
This followed from the analysis of the Green's function $G^{>}(x,t)$.
Beside a broad peak moving with $\bar{v}$, one observes a sharp peak
in time, moving along the trajectory $x=v_{F}t$, which is obviously
responsible for the contributions to $G^{>}(\epsilon,x)$ at large
energies. In addition, we confirm the assumption numerically. As we
are not able to perform the Fourier transformation analytically, instead
Fig. \ref{dispersionrelation} shows the numerical evaluation of the
function $e^{-F(x)}$ and the direct Fourier transform of the Green's
function $G^{>}(x,t)$ resulting from the bosonization in the limit
$\epsilon\rightarrow\infty$. Obviously, those turn out to be identical.
Therefore we conclude that the semiclassical approach is becoming
exact in the limit of high-energy electrons. We emphasize that, apart
from the transparent interpretation of the decoherence of high energetic
electrons via the semiclassical method, even the observation that
the Fourier transform of the Green's function at high energies turns
out to be identical to $[G^{>}(x,t)/G_{0}^{>}(x,t)]G_{0}^{>}(\epsilon,x)$$ $
with $x=v_{F}t$ is non-trivial and serves as a starting point for
the analytical discussion of the decoherence in this regime. 

Having established the connection between the semiclassical approach
and the full bosonization solution, we now turn to the properties
of the symmetrized potential spectrum. For this, first we focus on
the zero temperature case $T=0$ (see Fig.~\ref{dispersionrelation}).
At high frequencies, we obtain a singularity $\langle\{\hat{V},\hat{V}\}\rangle_{\omega}^{T=0}\propto1/\sqrt{\omega_{{\rm {max}}}-|\omega|}$
at the cutoff frequency $\omega_{{\rm max}}={\rm max}(\omega(q)-v_{F}q)$,
which is the maximum frequency in the Galilei-transformed plasmon
dispersion relation. Such a maximum frequency arises due to the momentum
cut-off in the dispersion relation (which results from the finite
range of the interactions). Due to this smooth momentum cut-off, the
velocity of plasmons in the limit of large momenta is identical to
the Fermi velocity $v_{F}$ of the electron. By transforming the potential
to the moving frame, it gets tilted (see Fig.\ref{dispersionrelation}).
As the velocity of the plasmons in the limit of large momenta is identical
to the velocity of the moving frame, the effective dispersion relation
shows a maximum. The singularity in the spectrum arises due to the
fact that $\omega(q)\approx\omega_{{\rm {max}}}+\omega''(q-q^{\ast})^{2}/2$
in the vicinity of $q^{*}$, where $\omega(q^{*})=\omega_{{\rm max}}$.
Note that an interaction potential $U_{q}$ with a non-monotonous
decay in $q$ may give rise to several such singularities, corresponding
to the local maxima of $\omega(q)-v_{F}q$. 

At low frequencies $\omega\ll v_{F}q_{c}$, the spectrum increases
linearly in $\omega$, corresponding to {}``Ohmic'' noise, which
is ubiquitous in various other physical contexts \cite{2000_Weiss_QuantumDissipativeSystems}.
For interaction potentials that are smooth in real space (i.e. where
all the moments of $\left|U_{q}\right|$ are finite), we find that
the leading low-frequency behaviour is determined solely by the contribution
to Eq.~(\ref{eq:VVomega}) stemming from small $q$. The result is
(here for $\alpha>0$):

\begin{equation}
\langle\{\hat{V},\hat{V}\}\rangle_{\omega}^{T=0}=\frac{U_{q\rightarrow0}^{2}}{(\bar{v}-v_{F})^{2}}\frac{|\omega|}{2\pi}=2\pi|\omega|.\label{eq:VVom}\end{equation}
Most remarkably, the dimensionless prefactor (the slope) of the noise
spectrum turns out to be completely independent of the coupling strength
$\alpha$, which drops out. This is in contrast to the typically studied
non-chiral Luttinger liquids, where an Ohmic spectrum has been found
with an interaction-dependent prefactor \cite{2005_LeHur_ElectronFractionalization}.
As a direct consequence, the electron's Green's function shows a universal
power-law decay at long distances, as we will discuss in more detail
in the next section.

\subsection{Universal dephasing for high-energy electrons}

\begin{figure}
\includegraphics[width=1\columnwidth]{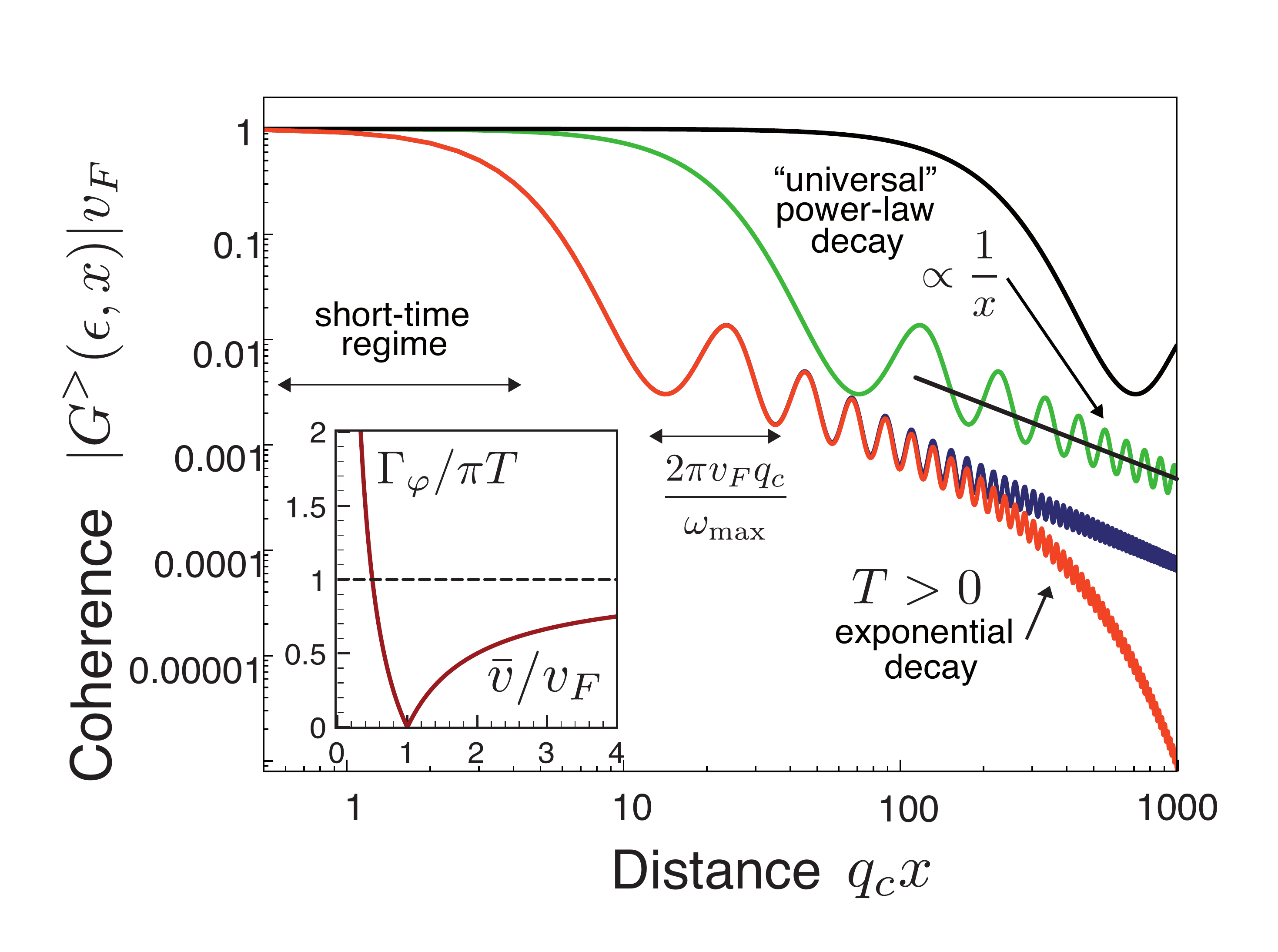}

\caption{\label{fig:UPLD}Coherence of an electron propagating at high energies
in an interacting chiral system, as a function of propagation distance
for various values of interaction strength $\alpha$: $2\pi\alpha=0.1$
(solid black line), $2\pi\alpha=1.0$ (solid green line) and $2\pi\alpha=5.0$
(solid red and blue). The potential is taken as $U_{q}=U_{0}e^{-|q/q_{c}|}$
and temperature is zero except for the red line where $T/q_{c}v_{F}=0.005$.
The non-interacting case would give $v_{F}\left|G^{>}(\epsilon,x)\right|\equiv1$.
The long-distance decay is universally given by $\propto1/x$, independent
of interaction strength. Note that for decreasing coupling strength
the asymptotic power-law decay sets in for increasingly larger propagation
distances. At finite temperatures, this power-law decay turns into
an exponential decay for large $x$ with a decay rate $\Gamma_{\varphi}$
depending on interaction strength (inset). }

\end{figure}
We insert (\ref{eq:VVom}) into the long-distance limit of Eq.~(\ref{eq:ReF}):\[
{\rm Re}[F(x)]=2\lim_{x\rightarrow\infty}\int_{0}^{+\infty}\frac{d\omega}{2\pi}\frac{\sin^{2}(\omega x/2v_{F})}{\omega^{2}}\left\langle \left\{ \hat{V},\hat{V}\right\} \right\rangle _{\omega}^{T=0}\]
\begin{equation}
\approx{\rm const}+\underbrace{\int_{v_{F}/x}^{\omega_{c}}d\omega\,\frac{1}{\omega}}_{\ln(x\omega_{c}/v_{F})}+2\int_{\omega_{c}}^{\omega_{{\rm max}}}\frac{d\omega}{2\pi}\frac{\sin^{2}(\omega x/2v_{F})}{\omega^{2}}\left\langle \left\{ \hat{V},\hat{V}\right\} \right\rangle _{\omega}^{T=0}\label{eq:Exp_SC}\end{equation}
where the choice of $\omega_{c}\ll\omega_{{\rm max}}$ is arbitrary
(but related to the constant), and $|G^{>}(\epsilon\rightarrow\infty,x)|v_{F}=e^{-{\rm Re}[F(x)]}$.
As a consequence of the logarithmic contribution to the exponent,
the leading asymptotic behaviour of the Green's function is a power-law
decay with an interaction-independent exponent $1$:

\begin{equation}
\left|G^{>}(\epsilon,x)\right|\propto\frac{1}{x}\,.\label{eq:PLD}\end{equation}
We recall the fact that Eq.$\,$(\ref{eq:PLD}) refers to the Green's
function in energy-coordinate space, as a function of propagation
distance $x$ at large energies. Having shown previously that the
current and thus the visibility only depend on this function (see
Eqs. (\ref{eq:I_coh},\ref{eq:visi_compact})), we may denote it the
'coherence' of the electron. Therefore, Eq.$\,$ (\ref{eq:PLD}) shows
that the coherence itself displays a universal power-law decay. Please
observe that in the absence of interactions $\left|G^{>}(\epsilon,x)\right|$
would be constant, so the decay indeed implies interaction-induced
decoherence. The surprising feature is the independence of the exponent
from the strength of the interactions (while the prefactor, not displayed
here, would indeed depend on the details of the potential).

In writing down Eq.~(\ref{eq:Exp_SC}), we have neglected the contributions
of large momenta in Eq.~(\ref{eq:VVomega}). These will lead to a
subleading correction to the power-law, which we discuss below. The
third term in Eq.~(\ref{eq:Exp_SC}) is responsible for oscillations
in the coherence $|G^{>}(\epsilon,x)|$, on top of the decay (see
below). Fig.~\ref{fig:UPLD} shows the decay of $G^{>}(\epsilon\rightarrow\infty,x)$
for different coupling constants $\alpha$. 

In order to understand how this generic result for the asymptotic
decay is compatible with the non-interacting limit ($\alpha=0$, where
$|G^{>}(\epsilon,x)|$ is constant), we have to discuss the range
of validity of the asymptotic behaviour. As the linear slope in the
effective spectrum $\left\langle \left\{ \hat{V},\hat{V}\right\} \right\rangle _{\omega>0}$
applies only at $|\omega|\ll\omega_{{\rm max}}$, we must certainly
require $\omega_{{\rm max}}x/v_{F}\gg1$. Since $\omega_{{\rm max}}$
vanishes linearly with $\alpha$, the limiting regime is reached at
ever larger values of $x$ when the interaction strength is reduced.
This shift of the range of validity is shown in Fig.~\ref{fig:UPLD}. 

The oscillatory modulation is due to the square root singularity at
$\omega\rightarrow\omega_{{\rm max}}$ in $\left\langle \left\{ \hat{V},\hat{V}\right\} \right\rangle _{\omega>0}$.
Its amplitude depends on the interaction strength $|\alpha|$ but
vanishes at long distances (see appendix \ref{sub:Amplitude-of-the}):\begin{equation}
F(x)\approx{\rm const}+\ln(\omega_{c}x/v_{F})-C\cdot\frac{\sin(\omega_{{\rm max}}x/v_{F}+\pi/4)}{\sqrt{2\pi|\alpha|q_{c}x}}\label{eq:}\end{equation}
where $C$ is a numerical constant, which solely depends on the explicit
form of the potential. 

We now discuss the deviations from the leading low-frequency behaviour
of $\left\langle \left\{ \hat{V},\hat{V}\right\} \right\rangle _{\omega}$.
Taking into account the contributions resulting from large $q$ in
Eq.~(\ref{eq:VVomega}), for smooth potentials like $U_{q}=U_{0}e^{-(|q|/q_{c})^{s}}$
the correction is $\left\langle \left\{ \hat{V},\hat{V}\right\} \right\rangle _{\omega}^{{\rm sub}}=2\pi\omega/(s\ln(\left|\alpha\right|v_{F}q_{c}/\omega))$
(\ref{sub:Sub-leading-corrections-for}). In real space this expression
translates into a correction $ $$\frac{1}{s}\ln(\ln(|\alpha|{}_{}q_{c}x))$
to the decay function $F(t=x/v_{F})$ (\ref{sub:Sub-leading-correction-in}). 

At finite temperature $T\neq0$, the long-time limit is given by an
exponential decay $|G^{>}(x,\epsilon)|\propto\exp[-\Gamma_{\varphi}x/v_{F}]$,
with a decay rate 

\begin{equation}
\Gamma_{\varphi}=\pi T\left|1-\frac{v_{F}}{\bar{v}}\right|=\pi T|1+\alpha^{-1}|^{-1}\,.\end{equation}
This follows from the long-time limit of Eq.~(\ref{eq:ReF}) together
with Eq.~(\ref{eq:VV_Sym_inv_sign}):\begin{equation}
{\rm Re}[F(x)]\approx2\int_{0}^{\infty}dq\,\frac{\sin^{2}(q|U_{q}|x/(4\pi v_{F}))}{q}\coth\left(\frac{|\omega(q\rightarrow0)|}{2T}\right).\label{eq:}\end{equation}
Using the identity $\lim_{a\rightarrow\infty}\frac{\sin^{2}(ax)}{ax^{2}}=\pi\delta(x)$,
we get ${\rm Re}[F(x)]=\pi T|1+\alpha^{-1}|^{-1}x$, from which we
obtain the aforementioned decay rate $\Gamma_{\varphi}$. In Fig.~\ref{fig:UPLD}
the decay rate is shown as a function of the coupling $\alpha$. For
small $\alpha$, this rate vanishes as $\Gamma_{\varphi}=\pi T|\alpha|$,
i.e. it is non-analytic in $U_{0}\propto\alpha$. Such dephasing rates
proportional to $T$ have also been found in non-chiral Luttinger
liquids \cite{2005_Mirlin_DephasingLLweakLoc,2005_LeHur_ElectronFractionalization,2006_LeHur_LifetimeLL}.
At large repulsive coupling, $U_{0}\rightarrow+\infty$, we have the
universal result $\Gamma_{\varphi}\rightarrow\pi T$. For attractive
interaction, $\Gamma_{\varphi}$ diverges upon approaching the instability
at $\alpha\rightarrow-1$, where $\bar{v}\rightarrow0$ and where
the resulting low-frequency modes are thermally strongly excited.

We note that this behaviour is somewhat surprising when compared to
other problems of dephasing. When considering pure dephasing of a
two-level system by an Ohmic bath, a power-law decay $t^{-\gamma}$
at $T=0$, with an exponent $\gamma$ set by the coupling, automatically
implies an exponential decay at a rate $\Gamma_{\varphi}=\pi\gamma T$
at finite temperatures. This follows from the fluctuation-dissipation
theorem (FDT) which turns the $T=0$ Ohmic spectrum into a white-noise
spectrum with a weight proportional to $T$. 

However, in the present case, we have to take into account the Galileo
transformation, which turns the laboratory-frame temperature $T$
into an effective temperature $T_{{\rm eff}}$ in the frame moving
along with the particle at speed $v_{F}$. In order to establish the
FDT for the effective potential fluctuations in the electron frame
of reference we therefore have to define the effective temperature
$T_{{\rm eff}}$ by demanding for $\omega\downarrow0$ \begin{eqnarray}
 & \left\langle \left\{ \hat{V},\hat{V}\right\} \right\rangle _{\omega}= & \int_{0}^{\infty}\frac{dq}{2\pi}\,|U_{q}|^{2}q\coth\left(\frac{\omega(q)}{2T}\right)\delta(\omega-\frac{|U_{q}q|}{2\pi})\nonumber \\
 & \equiv & \coth\left(\frac{\omega}{2T_{{\rm eff}}}\right)\left\langle \left[\hat{V},\hat{V}\right]\right\rangle _{\omega}.\label{eq:T_eff}\end{eqnarray}
In the low-frequency limit we get $ $$\left\langle \left\{ \hat{V},\hat{V}\right\} \right\rangle _{\omega}=2\pi\omega\cdot\coth\left(\omega\left|1+\alpha^{-1}\right|/(2T)\right)$.
Together with Eq.~(\ref{eq:VV_commutator}), this yields the effective
temperature $T_{{\rm eff}}\equiv T\left|1+\alpha^{-1}\right|^{-1}$.
In the moving frame the frequencies are reduced by $q_{c}v_{F}$ and
therefore the effective temperature is also smaller.Only for large
repulsive interactions ($\bar{v}\gg v_{F}$), the transformation does
not matter. Therefore, in this limit $T_{{\rm eff}}=T$ and the universal,
coupling-independent power-law for $T=0$ turns into a universal decay
rate at finite temperatures. 

Finally, turn briefly to the question of observing these features
in experiments. We note that the universal power-law decay of the
Green's function should be observable in principle in the Mach-Zehnder
interferometer setup, as it directly translates into a decay of the
visibility itself. To obtain a numerical estimate, we assume a screened
Coulomb potential with screening length $q_{c}^{-1}\sim10^{-7}m$
and a finite channel width $b\sim10^{-7}m$. Following \cite{2007_ChalkerGefen_MZ}
the edge state velocity can be assumed to be: $v_{F}\sim(10^{4}-10^{5})\frac{m}{s}$.
Then the dimensionless coupling constant $\alpha$ is of the order
$1$. To reach the high energy limit, the applied bias voltage $V$
has to fulfill: $q_{e}V\gg q_{c}v_{F}$. For $v_{F}=10^{4}\frac{m}{s}$,
it turns out that one has to apply $V\sim10^{2}\mu V$ which is in
the range of the bias voltages typically applied in experiments (in
\cite{2006_01_Neder_VisibilityOscillations} $V\sim10^{1}\mu V$).
We mention that the long-distance limit should be reached for $x\gg v_{F}/\omega_{{\rm max}}\sim10^{-1}\mu m$,
which is shorter than the typical interferometer arm length (e.g,
in \cite{2006_01_Neder_VisibilityOscillations} $x_{1,2}\sim10^{1}\mu m$).
Thus there is some hope that the high-energy as well as the long-distance
limit is accessible in the experiment. However, we note that the magnitude
of the Green's function $G^{>}(\epsilon,x)$ gets suppressed strongly
when reaching the long-distance limit, i.e. $x\gg v_{F}/\omega_{{\rm max}}$.
For $\alpha=1$, the magnitude of $|G^{>}(\epsilon,x)|$ yields: $|G^{>}(\epsilon,x)|\sim10^{^{-2}}$
(see Fig.\ref{fig:UPLD}). With help of Eq.$\,$(\ref{eq:visi_compact})
this translates into a visibility of the order $v_{I}\sim10^{-4}$$ $
(for this see also section \ref{sec:Visibility-and-Current}). Therefore,
the direct measurement of the power-law dependence of the visibility
for large bias voltages seems to be a challenging task, unless one
finds a way to optimize further the relevant parameters.

\section{Keldysh perturbation theory\label{sub:Low-energy--}}

\subsection{General remarks}

\begin{figure}
\includegraphics[width=1\columnwidth]{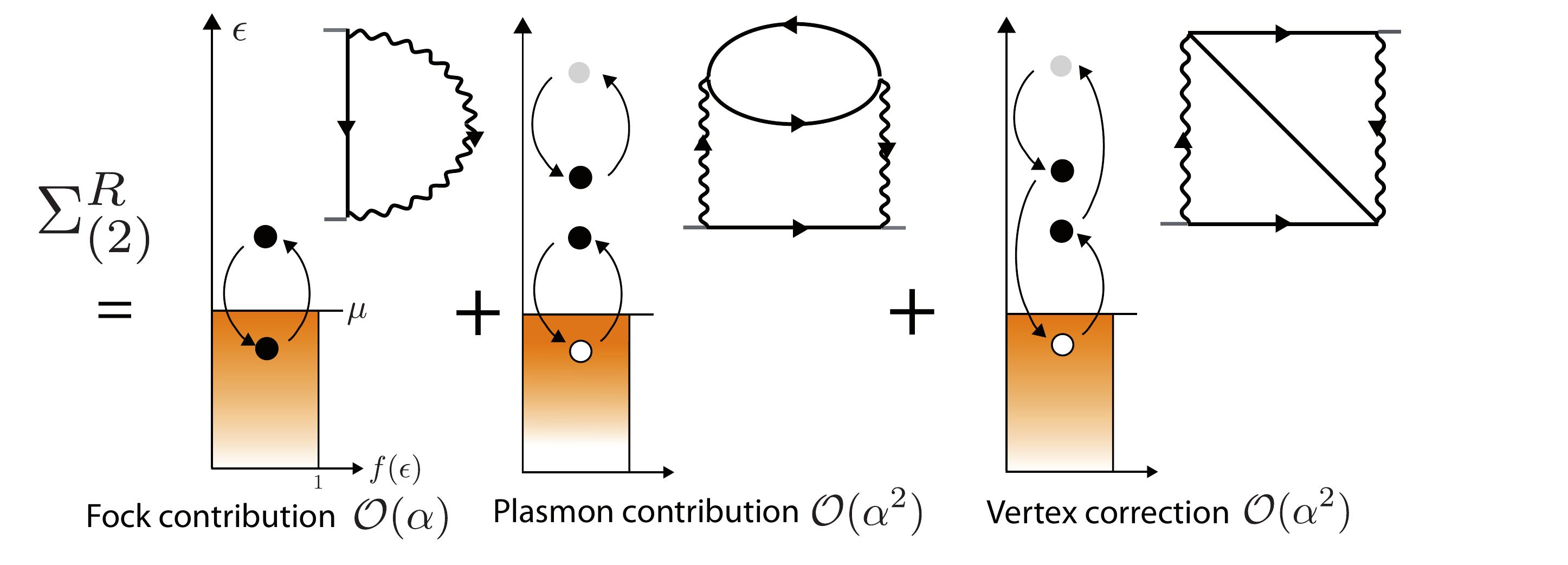}\caption{The relevant processes which contribute to the self-energy up to second
order in the coupling. The wiggly lines indicate the interaction.
Note that the plasmonic diagram and the vertex correction differ only
by an additional exchange process. That produces a minus sign, such
that the diagrams tend to cancel each other at low momenta. \label{KeldyshProc}}

\end{figure}

The semiclassical method we introduced in the preceding section provides
us with an intuitive picture for the case of electrons flying high
above the Fermi sea. To complete the analysis, in the present section
we employ perturbation theory to discuss the behaviour at short propagation
distances and weak coupling for all energies. This includes the low
energy regime, where the influence of the Fermi edge becomes important.
We will apply Keldysh (i.e. nonequilibrium) perturbation theory up
to second order in the interaction strength $\alpha$. The main outcome
of the perturbation theory is that the tunnel density of states is
affected by renormalization effects while the decay of the GF in the
close vicinity of the Fermi edge is suppressed. We will find that
the suppression of decoherence is brought about by a cancellation
between two second order diagrams near the Fermi edge. 

During this subsection \textcolor{black}{we fix the notation}, following
the review \cite{1986_RammerSmith}\textcolor{black}{.} Using the
Keldysh time leads to an additional matrix structure of the GF which
reflects the fact that one has to differentiate between points in
time which lie on the backward or the forward branch of the Keldysh
contour. Beyond this additional structure, all the known Feynman rules
remain exactly the same. After performing the rotation in Keldysh
space \cite{1986_RammerSmith}, the representation of the matrix
GF $G$ and the related matrix self-energy is given by \begin{equation}
G\equiv\left[\begin{array}{cc}
G^{R} & G^{K}\\
0 & G^{A}\end{array}\right]\quad\quad\Sigma\equiv\left[\begin{array}{cc}
\Sigma^{R} & \Sigma^{K}\\
0 & \Sigma^{A}\end{array}\right]\label{eq:}\end{equation}
where we introduce the Keldysh GF, $G^{K}(x,t,x',t')\equiv-i\langle[\hat{\Psi}(x,t),\hat{\Psi}^{\dagger}(x',t')]\rangle$.
First we will derive an expression for the retarded self-energy $\Sigma^{R}(\epsilon.k)$,
which will be used to calculate $G^{R}$ that can be related to the
single particle propagator $G^{>}$ in equilibrium. Starting from
the matrix Dyson equation $G(\epsilon,k)=G_{0}(\epsilon,k)+G_{0}(\epsilon,k)\cdot\Sigma(\epsilon,k)\cdot G(\epsilon,k)$,
one finds that the retarded Green's function only depends on the retarded
self-energy:\textbf{ }\begin{equation}
G^{R}(\epsilon,k)=\frac{1}{[\epsilon-\epsilon_{0}(k)+i0^{+}]-\Sigma^{R}(\epsilon,k)}.\label{eq:G_ret&Self}\end{equation}
In the following we calculate the diagrams up to second order for
a linearized dispersion\textbf{ }relation, but for finite temperature
and for an arbitrary interaction potential. In the end we compare
the results of the perturbation theory with the results of the bosonization
technique. The relevant processes are shown in Fig.\ref{KeldyshProc}.
There are two second order diagrams, which can be identified as the
interaction with a plasmonic excitation and a corresponding diagram
containing an additional exchange process (that can be viewed as a
vertex correction diagram). The crucial point is that the vertex correction
counteracts the plasmonic processes in the vicinity of the Fermi edge,
leading to a suppression of the decay of the GF.

\subsection{Evaluation of the diagrams}

The starting point of the calculation is the evaluation of the the
unperturbed electronic propagator matrix $G_{0}$. In addition to
the usual retarded and advanced GFs of free electrons, $G_{0}^{R/A}(\omega,k)=\frac{1}{\omega-v_{F}k\pm i0^{+}}$,
the Keldysh propagator is given by $G_{0}^{K}(\omega,k)=-2\pi i\tanh(\frac{\beta\omega}{2})\cdot\delta(\omega-v_{F}k)$.
In contrast to the advanced and retarded GFs, the Keldysh propagator
contains information about the electronic spectrum as well as about
the occupation of those states. Therefore at this point one could
introduce arbitrary non-equilibrium states which is the main advantage
of working on the Keldysh contour. However, as we are describing channels
which are only weakly tunnel-coupled to each other, we will calculate
equilibrium Green's functions. 

As in second order we explicitly include interactions with free plasmons,
we also derive their propagators here. To this end we identify the
bosonic field with the potential $\hat{V}(x,t)=\int dx'\, U(x-x')\hat{\rho}(x',t)$.
This is identical to the potential we introduced in the semiclassical
description {[}although in the latter case we specialized to $x\equiv v_{F}t${]}.
The bosonic propagators are defined as $D^{R/A}(x,t)=\mp i\theta(\pm t)\langle[\hat{V}(x,t),\hat{V}(0,0)]\rangle$
and $D^{K}(x,t)=-i\langle\{\hat{V}(x,t),\hat{V}(0,0)\}\rangle$, respectively.
A straightforward calculation yields for the retarded and advanced
propagators

\begin{equation}
D^{R/A}(\omega,q)=U_{q}^{2}\cdot\int\frac{dk}{2\pi}\frac{f(k)-f(q+k)}{(\omega+\epsilon_{0}(k)-\epsilon_{0}(k+q))\pm i0^{+}}=U_{q}^{2}\frac{q}{2\pi}\cdot G^{R/A}(\omega,q).\label{eq:plamProp}\end{equation}
As we assume the system to be in equilibrium, we use the FDT to obtain
the plasmonic Keldysh propagator: \begin{equation}
D^{K}(\omega,q)=2i\coth(\beta\omega/2){\rm Im}[D^{R}(\omega,q)].\label{eq:}\end{equation}
Now we can proceed calculating the various contributions to the self-energy
up to $\mathcal{O}(\alpha^{2})$. Considering all the possible Feynman
diagrams, we are left with the first order Hartree-Fock diagrams and
in second order with the plasmon diagram and the vertex correction
(see Fig.\ref{KeldyshProc}). Thus we can express the self-energy
as $\Sigma_{(2)}^{R}=\Sigma_{{\rm Hartree}}+\Sigma_{{\rm Fock}}+\Sigma_{{\rm Plasmon}}+\Sigma_{{\rm Vertex}}$.
Those can be evaluated according to the rules given in \cite{1986_RammerSmith}.
We use the equal-time interaction propagators $U^{R/A}(q)\equiv U_{q}$
and set $U^{K}\equiv0$, as usual.

\paragraph{First-order contributions: Hartree-Fock diagrams}

The Hartree diagram yields a global energy renormalization $\Delta E_{{\rm Hartree}}=U(q=0)\bar{\rho}$.
In the framework of the Luttinger model the electron density diverges
as there is no lower boundary of the electron spectrum. However, formally
one can include the energy shift into the definition of the chemical
potential (see subsection \ref{sub:Hamiltonian}). In the following
we omit the Hartree contribution.

\begin{figure}[t]
\includegraphics[width=1\columnwidth]{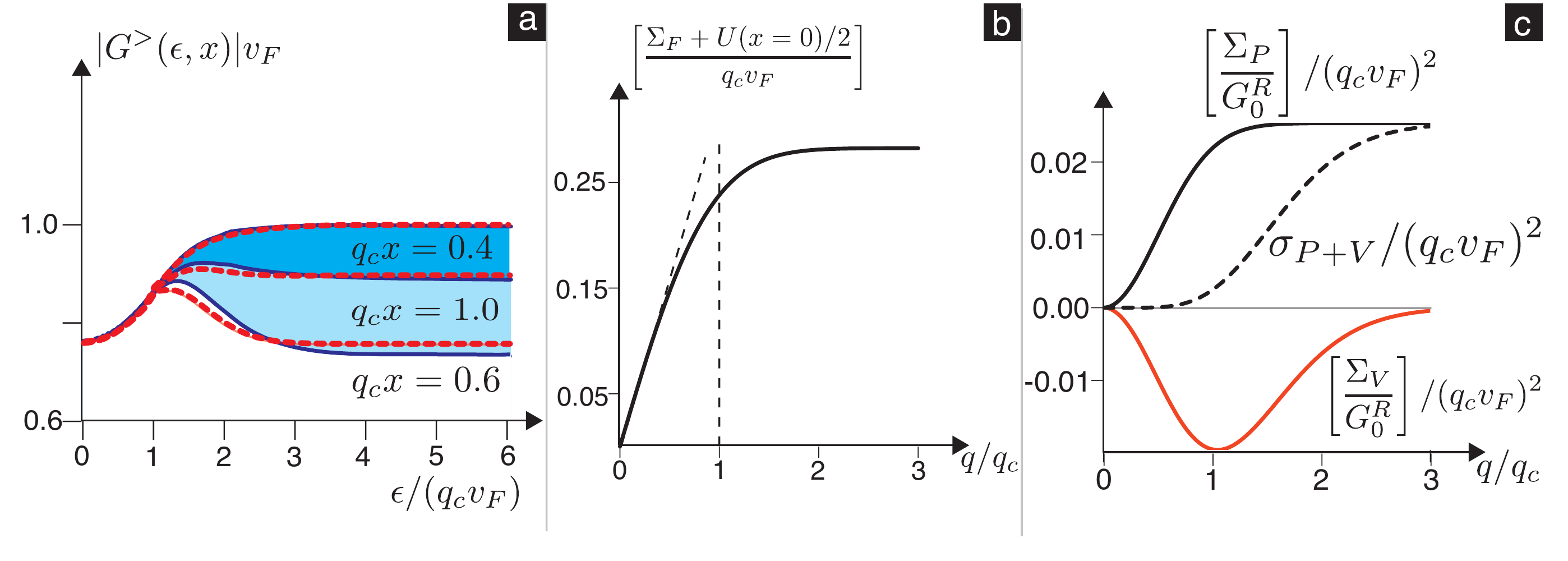}\caption{(a) The GF from bosonization (solid blue lines) vs. the results from
second-order Keldysh pertubation theory (dashed red lines) for different
propagation distances as a function of energy. Note the fairly good
agreement. (b) Fock contribution to the self-energy $\Sigma_{F}^{R}(k)$,
which starts with a linear slope (corresponding to the electron velocity
renormalization) and then saturates for large momenta. (c) Total second
order contribution the self-energy (dashed, black line); plasmonic
(solid black line) and vertex correction (lower, red line) separately.
The cancellation for small momenta is discussed in detail in the text
(in all figures $U_{q}=U_{0}\exp(-(q/q_{c})^{2})$ and $2\pi\alpha=U_{0}/v_{F}=2$\foreignlanguage{german}{)
\label{fig:Keldysh/Boso}}}

\end{figure}

The self-energy contribution due to the Fock diagrams is \begin{equation}
\Sigma_{F}^{R}(\omega,k)=-\frac{1}{2\pi}\int_{}^{}dq\, U_{q}f(k-q),\label{eq:hFcontri}\end{equation}
which for zero temperature yields $\Sigma_{F,T\equiv0}^{R}(k)=-\frac{1}{2}U(x=0)+\frac{1}{2\pi}\int_{0}^{|k|}dq\, U_{q}$$ $
{[}see Fig.\ref{fig:Keldysh/Boso}b{]}. 

The resulting $k$-dependent energy shift describes, in particular,
the renormalization of the eletron velocity near the Fermi edge. This
also affects the tunneling density of states, leading to a suppression
(for repulsive interactions, $\alpha>0$) or enhancement ($\alpha<0$).
This can be seen in Fig.\ref{gexfig}. Again, the constant shift can
be incorporated in the definition of the chemical potential.

\paragraph{Second-order contributions: Plasmonic excitations and vertex correction}

The electron's coherence decays by interacting with the plasmons,
i.e. the density fluctuations of the other electrons. The plasmon
diagram (see Fig.~\ref{KeldyshProc}) represents one of the contributions
describing this physics. It yields

\begin{equation}
\Sigma_{P}^{R}(\epsilon,k)=\frac{i}{2}\int(dq)\int(d\omega)\left[G_{0}^{R}(\epsilon-\omega,k-q)\cdot D_{0}^{K}(\omega,q)+G_{0}^{K}(\epsilon-\omega,k-q)\cdot D_{0}^{R}(\omega,q)\right]\,,\label{eq:}\end{equation}
where the second term contains the Fermi function, which introduces
the effects of the Fermi edge on the coherence. Inserting the propagators,
the contribution can be written in a compact form

\begin{equation}
\Sigma_{P}^{R}(\epsilon,k)=\frac{G_{0}^{R}(\epsilon,k)}{8\pi^{2}}\cdot\int_{-\infty}^{\infty}dq\quad U_{q}^{2}q\cdot\left[\coth(\beta\hbar v_{F}q/2)+\tanh(\beta\hbar v_{F}(k-q)/2)\right],\label{eq:plasmonself}\end{equation}
which for $T=0$ reduces to $\Sigma_{P}^{R}(\epsilon,k)=G_{0}^{R}(\epsilon,k)\cdot\frac{1}{4\pi^{2}}\int_{0}^{|k|}dq\, U_{q}^{2}q$.
Thus, at $T=0$ this contribution vanishes for $k\rightarrow0$. We
note in passing that the structure {}``$\coth+\tanh$'' generically
occurs in discussions of dephasing, where it describes both the strength
of the thermal fluctuations and the influence of the Fermi function,
i.e. the physics of Pauli blocking \cite{1983_FukuyamaAbrahams,2004_10_Marquardt_MZQB_PRL,2006_04_MZQB_Long,2006_04_DecoherenceReview,2005_10_WeaklocDecoherenceOne,2005_10_WeaklocDecoherenceTwo}.
In the limit of high energies, the result of Eq.~(\ref{eq:plasmonself})
can be rewritten in terms of the potential fluctuations at the particle
position, as discussed in the preceding section. Specifically, we
have $\lim_{k\rightarrow\infty}\Sigma_{P,T\equiv0}^{R}=G_{0}^{R}\cdot\langle\hat{V}(x=0,t=0)^{2}\rangle$.
For a plot of the function $\Sigma_{P,T=0}^{R}$, see Fig.\ref{fig:Keldysh/Boso}c. 

Finally we derive the vertex correction, mentioned above, which after
a rather lengthy calculation yields (see appendix \ref{sub:Vertex})

\[
\Sigma_{V}^{R}(\epsilon,k)=\left[\frac{G_{0}^{R}(\epsilon,k)}{16\pi^{2}}\right]\int_{-\infty}^{\infty}dq_{1}\int_{-\infty}^{\infty}dq_{2}U_{q_{1}}U_{q_{2}}\]
\[
\times\left[\tanh(\frac{\beta v_{F}(k-q_{1}-q_{2})}{2}))\cdot\left[\tanh(\frac{\beta v_{F}((k-q_{1})}{2})+\tanh(\frac{\beta v_{F}((k-q_{2})}{2}))\right]-\right.\]
\begin{equation}
\left.-\tanh(\frac{\beta v_{F}((k-q_{2})}{2})\ \tanh(\frac{\beta v_{F}((k-q_{1})}{2})-1\right].\label{eq:}\end{equation}
This expression simplifies for $T=0$ to $ $$\Sigma_{V,T=0}^{R}=-G_{0}^{R}\cdot\frac{1}{4\pi^{2}}\int_{0}^{|k|}dq_{1}\int_{|k|-q_{1}}^{|k|}dq_{2}U_{q_{1}}U_{q_{2}}$.
The contribution from the vertex correction as well as the total second
order correction to the self-energy $\sigma_{{\rm P+V}}^{R}\equiv\Sigma_{P}^{R}+\Sigma_{V}^{R}$
are shown in Fig.\ref{fig:Keldysh/Boso}. The crucial feature is that
up to second order in momentum $k$ the plasmon diagram and the vertex
correction cancel exactly against each other, whereas for high momenta
{[}$k\gg q_{c}$ {]} only the plasmon contribution remains while the
vertex correction tends to zero. In summary, when calculating up to
second order in the coupling, we already see that the dephasing is
suppressed in the vicinity of the Fermi edge. As the comparison with
the exact bosonization solution shows, this conclusion holds true
qualitatively to all orders. We can the preceding expressions to evaluate
the retarded Green's function $G^{R}$ and from this the propagator
$G^{>}$. The final results has evaluated numerically. The results
are illustrated in Fig.\ref{fig:Keldysh/Boso}. Expanding the Green's
function for small propagation distance $x\ll v_{F}\left(2\sqrt{\langle\hat{V}(0)^{2}\rangle}\right)^{-1}$,
yields:\begin{equation}
|G^{>}(\epsilon\rightarrow\infty,x)|v_{F}\approx1-\frac{1}{2}\langle\hat{V}(0)^{2}\rangle(\frac{x}{v_{F}})^{2}+\ldots,\end{equation}
which coincides with the expanded exact result. The good agreement
of the bosonization result and the Keldysh perturbation theory for
small $|\alpha|$ is shown in Fig.~\ref{fig:Keldysh/Boso}, as well
as in Fig.~\ref{dispersionrelation}. However, for large $x$ the
perturbation theory fails.

For a detailed study of chiral interacting electrons of the spectrum
in $(\epsilon-k)$-space, starting from the bosonization result, we
refer the reader to \cite{1993_MedenSchoenhammer_SpectrumLL}. 

\subsection{Summary of the Keldysh perturbation theory}

To summarize, the perturbation theory shows that there are two different
energy regimes: In general, the GF decays as a function of propagation
distance due to the interaction with the density fluctuations. This
is particularly pronounced at high energies $\epsilon\gg v_{F}q_{c}$,
where we have also shown that the Keldysh result and the semiclassical
(or bosonization) approach coincide at short distances. For low energies
$\epsilon\ll v_{F}q_{c}$, the decay is suppressed. On the other hand,
at low energies the Fermi velocity is renormalized due to virtual
processes, leading to a modification of the tunneling density of states.
It is important to note that these two different energy regimes only
emerge since we are dealing with interaction potentials of finite
range.

\section{Visibility and Current\label{sec:Visibility-and-Current}}

\begin{figure}
\includegraphics[width=1\columnwidth]{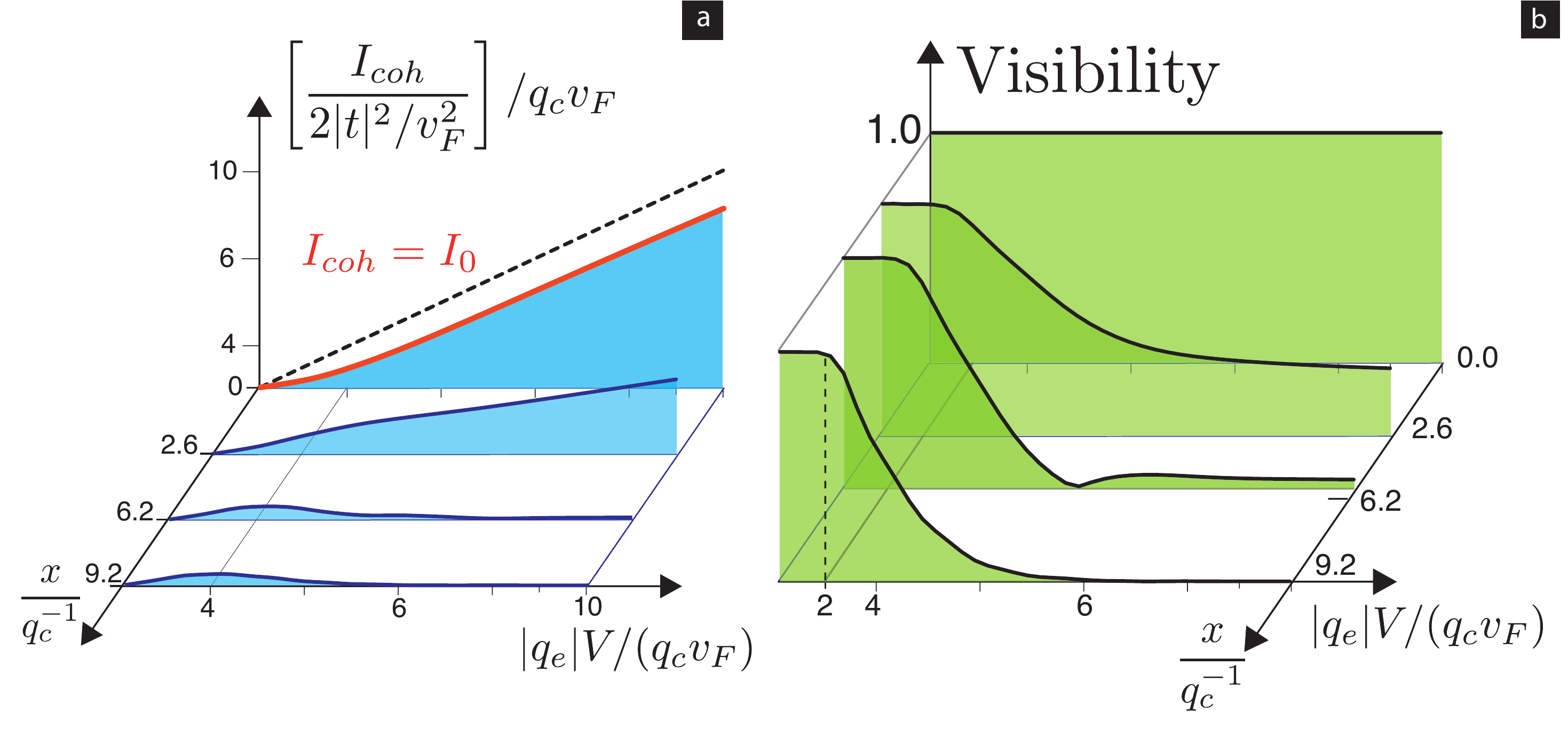}

\caption{a) The amplitude of the coherent (flux-dependent) part of the current
$I_{{\rm coh}}$ through the interferometer for different arm lengths
$x$. The curve for $x\equiv0$ is equal to the flux-independent current
$I_{{\rm coh}}(x\equiv0,V)=I_{0}(V)$ (red line), which implies that
the interference contrast is perfect at zero armlength. \protect \\
b) The visibility $v_{I}$ as a function of bias voltage $V$ for
various arm-lengths $x_{1}=x_{2}=x$. The red line denotes the semiclassical
calculation. The small deviations from the bosonization result vanish
completly for larger $V$. The plot is done for $U_{q}=U_{0}e^{-(q/q_{c})^{2}}$
with $\frac{U_{0}}{v_{F}}=2\pi\alpha=3$.\label{Visibility}}

\end{figure}

As mentioned in the beginning, the results for the GF $G^{>}(\epsilon,x)$
we worked out in the foregoing section can be applied directly to
the evaluation of the current and the visibility (Fig.\ref{Visibility}). 

Fig.\ref{Visibility}a shows the current through the interferometer
as a function of voltage for different arm lengths. Here we restrict
the considerations to the symmetric case, i.e. $x_{1}=x_{2}$. For
$x=0$ the coherent part of the current obviously is identical the
flux-independent part, which implies a perfect visibility (at $T=V=0$).
The suppression of the current at small voltages is due to the velocity
renormalization which lowers the tunnel density (for repulsive interactions).
However, as the change in the tunnel density influences the classical
and the coherent part in the same way, it does not show up in the
visibility at all, i.e. $v_{I}(V,x=0)\equiv1$ (see Fig.\ref{Visibility}b). 

The behaviour of the coherence of the GF for high energies discussed
above (using the semiclassical approach) can be transferred directly
to the discussion of the visibility. Therefore, in the limit of high
voltages $V$ and $T=0$ the visibility is determined by the factor
$|G^{>}(\epsilon\rightarrow\infty,x)|^{2}$. This follows from the
fact that for higher voltages the contribution of the high-energy
electrons becomes dominant. It also implies that the visibility at
high voltages becomes voltage-independent. Correspondingly, the exponential
decay at finite temperature is transferred to the visibility $v_{I}$
as well. For $x\neq0$ the dephasing reduces the coherent (flux-dependent)
part of the current which leads to a decrease of the visibility. At
small voltages $|q_{e}V|\ll q_{c}v_{F}$ the visibility decays only
very slowly with increasing interferometer length (see the discussion
of the Green's function). In the limit $V\rightarrow0$ the visibility
is approaching unity $v_{I}\rightarrow1$ , which is consistent with
the fact that in equilibrium and at zero temperature there is no dephasing.
In contrast to the Green's function itself, which shows oscillations
as a function of $\epsilon$, the visibility does not show pronounced
oscillations as a function of $V$.

\section{Conclusions}

In the present paper, we have studied dephasing by electron-electron
interactions in a ballistic interferometer. We have considered the
case of electrons moving inside ballistic, one-dimensional, chiral
channels, such as edge channels in the integer quantum Hall effect.
The interference contrast, in the limit of low transmission at the
beam splitters, can be expressed via the Green's function of the interacting
system. We have studied the decay of the electron's coherence as a
function of propagation distance, employing three different approaches:
The exact bosonization solution, a semiclassical approach that becomes
exact for high energies, and diagrammatic nonequilibrium (Keldysh)
perturbation theory. Our most important physical result is that at
high energies the decay of coherence at $T=0$ becomes a power-law
with a universal exponent, independent of interaction strength. We
have also shown that second-order perturbation theory compares well
with the exact solution at modest coupling strength and short distances.
This may be important for potential applications to more elaborate
setups that cannot be treated exactly any more.

\emph{Acknowledgements}. \textendash{} We thank J. Chalker, Y. Gefen,
V. Meden, B. Kubala, and O. Yevtushenko for interesting discussions
related to this work. Financial support by DIP, NIM, the Emmy-Noether
program and the SFB/TR 12 is gratefully acknowledged. \newpage{}

\appendix

\section{Semiclassical approach}

\subsection{\label{sub:Amplitude-of-the}Amplitude of the oscillations in $G^{>}(\epsilon,x)$
as a function of $x$ for $T=0$ }

The third term in Eq$\,$.$\,$(\ref{eq:Exp_SC}) is responsible for
the oscillations in $G^{>}(\epsilon\rightarrow\infty,x)$, while the
main contribution stems from the square-root singularity at $\omega=|\omega_{{\rm max}}|$.
This contribution yields:

\[
\delta F(x)=2\int_{|\omega_{{\rm max}}|-\delta}^{|\omega_{{\rm max}}|}(d\omega)\,\frac{\sin^{2}(\omega x/(2v_{F}))}{\omega^{2}}\cdot\left\langle \left\{ \hat{V},\hat{V}\right\} \right\rangle _{\omega\approx|\omega_{{\rm max}}|}\]
\begin{equation}
={\rm const.}-{\rm Re}\left[\int_{|\omega_{{\rm max}}|-\delta}^{|\omega_{{\rm max}}|}(d\omega)\,\frac{e^{i\omega\frac{x}{v_{F}}}}{\omega^{2}}\cdot\left\langle \left\{ \hat{V},\hat{V}\right\} \right\rangle _{\omega\approx|\omega_{{\rm max}}|}\right].\label{eq:deltaF}\end{equation}
From Eq.$\,$(\ref{eq:VV_Sym_inv_sign}) we obtain the $\omega$-dependence
of the correlator in the vicinity of the singularity\begin{equation}
\left\langle \left\{ \hat{V},\hat{V}\right\} \right\rangle _{\omega}\approx\frac{1}{2\pi}\frac{U^{2}(q_{0})q_{0}}{2\sqrt{|\xi|(|\omega_{{\rm max}}|-\omega)}}\label{eq:}\end{equation}
where $\omega(q^{\ast})={\rm max}(|\omega_{q}-v_{F}q|)$, $q_{0}\equiv q^{\ast}+\sqrt{\frac{|\omega_{max}|-\omega}{|\xi|}}$
and $\xi\equiv\frac{1}{2}\frac{d^{2}}{dq^{2}}\left(\frac{|U_{q}|q}{2\pi}\right)_{q=q^{\ast}}$.
Plugging in this result into Eq.$\,$(\ref{eq:deltaF}) yields \[
\delta F\approx-\frac{1}{8\pi^{2}}\frac{U^{2}(q^{\ast})q^{\ast}}{\omega_{{\rm max}}^{2}\sqrt{|\xi|}}\cdot{\rm Re}\left[\int_{|\omega_{{\rm max}}|-\delta}^{|\omega_{{\rm max}}|}d\omega\,\frac{e^{i\omega\frac{x}{v_{F}}}}{\sqrt{|\omega_{{\rm max}}|-\omega}}\right]\]
\begin{equation}
=-\frac{1}{8\pi^{2}}\frac{U^{2}(q^{\ast})q^{\ast}}{\omega_{{\rm max}}^{2}\sqrt{|\xi|}}\cdot Re\left[\int_{0}^{\delta\frac{x}{v_{F}}}d\nu\,\frac{e^{i\nu}}{\sqrt{\nu}}\cdot\frac{e^{i|\omega_{{\rm max}}|\frac{x}{v_{F}}}}{\sqrt{t}}\right].\label{eq:}\end{equation}
Evaluating the integral in the limit $x\rightarrow\infty$ gives\[
\delta F(x)\approx-\left[\frac{1}{8\pi^{2}}\frac{U^{2}(q^{\ast})q^{\ast}}{\omega_{{\rm max}}^{2}\sqrt{|\xi|}}\right]\cdot\sqrt{\frac{\pi}{t}}\cdot\sin(|\omega_{{\rm max}}|\frac{x}{v_{F}}+\pi/4)\]
\begin{equation}
=-C\cdot\frac{\sin(|\omega_{{\rm max}}|\frac{x}{v_{F}}+\pi/4)}{\sqrt{2\pi|\alpha|q_{c}x}}\label{eq:}\end{equation}
where $C$ denotes a numerical prefactor which depends only on the
form of the interaction potential, not on the interaction strength
$\alpha$.

\subsection{Sub-leading corrections for smooth potentials $U_{q}$ at $T=0$
\label{sub:Sub-leading-corrections-for}}

The subleading corrections result from the contributions of large
$q$ in Eq.$\,$(\ref{eq:VV_Sym_inv_sign}) (here $\omega>0$). We
start with the expression \begin{equation}
\left\langle \left\{ \hat{V},\hat{V}\right\} \right\rangle _{\omega}=\int_{0}^{\infty}dq\, U_{q}^{2}q\delta(\omega-\frac{q|U_{q}|}{2\pi}).\label{eq:}\end{equation}
The sub-leading correction to the low-frequency behaviour of $ $$\left\langle \{\hat{V},\hat{V}\}\right\rangle _{\omega}$
for a smooth potential $U_{q}=U_{0}e^{-|q/q_{c}|^{s}}$ is given by\begin{equation}
\left\langle \left\{ \hat{V},\hat{V}\right\} \right\rangle _{\omega}^{({\rm sub})}=\int_{0}^{\infty}dq\, U_{q}^{2}q\frac{\delta(q-q_{{\rm sub}}^{})}{\left|\frac{\partial}{\partial q}(|U_{q}|q)\right|_{q=q_{{\rm sub}}}},\label{eq:VV_sub}\end{equation}
where $q_{{\rm sub}}$ fulfills: $\frac{2\pi\omega}{|U_{0}|q_{c}}=\frac{q_{{\rm sub}}}{q_{c}}e^{-|q_{{\rm sub}}/q_{c}|^{s}}$.
Therefore, as $q\gg q_{c}$ for $\omega$ small enough we obtain for
$q_{{\rm sub}}$\begin{equation}
q_{{\rm sub}}=q_{c}\ln\left[\frac{q_{c}|U_{0}|}{2\pi\omega}\right]^{1/s}.\label{eq:}\end{equation}
Finally, the subleading corrections yields\begin{equation}
\left\langle \left\{ \hat{V},\hat{V}\right\} \right\rangle _{\omega}^{({\rm sub})}=\frac{2\pi\omega}{s\ln(\frac{|\alpha|v_{F}q_{c}}{\omega})},\label{eq:VV_sub2}\end{equation}
for $\omega\ll|\alpha|v_{F}q_{c}$.

\subsection{Sub-leading correction in time-domain\label{sub:Sub-leading-correction-in}
for $T=0$}

The subleading correction to the decay function $F(x)$ results from
the expression $F^{({\rm sub})}=\int_{-\infty}^{\infty}\frac{d\omega}{2\pi}\,\frac{\sin^{2}(\omega x/(2v_{F}))}{\omega^{2}}\cdot\left\langle \left\{ \hat{V},\hat{V}\right\} \right\rangle _{\omega}^{({\rm sub})}$.
With Eq.$\,$$\,$(\ref{eq:VV_sub2}) (omitting an additional constant)
it yields \begin{equation}
F^{({\rm sub})}(x)\approx\frac{2}{2}\int_{v_{F}/x}^{\omega_{c}}(d\omega)\,\frac{2\pi\omega}{\omega s\ln(\frac{\omega^{\ast}}{\omega})}=-\frac{1}{s}\int_{v_{F}/(\omega^{\ast}x)}^{\omega_{c}/\omega^{\ast}}d\omega\,\frac{1}{\omega\ln\omega},\label{eq:}\end{equation}
where for reasons of brevity we set $|\alpha|v_{F}q_{c}\equiv\omega^{\ast}$.
As we can choose the cutoff frequency $\omega_{c}$ such that $\omega_{c}/\omega^{\ast}<1$,
we can assume $0<\omega<1.$ This enables the substitution $\omega=e^{-x}$:\[
F^{({\rm sub})}(t)=-\frac{1}{s}\int_{\ln(\omega^{\ast}x/v_{F})}^{\ln(\omega^{\ast}/\omega_{c})}dx\,\frac{1}{x}=\]
\begin{equation}
=\frac{1}{s}\int_{\ln(\omega^{\ast}/\omega_{c})}^{\ln(\omega^{\ast}x/v_{F})}dx\,\frac{1}{x}=\frac{1}{s}\ln(\ln(|\alpha|q_{c}x))-{\rm const}.\label{eq:}\end{equation}

\section{Keldysh perturbation theory: Vertex correction\label{sub:Vertex}}

The vertex correction to the retarded self-energy has the structure
(compare to the review of \cite{1986_RammerSmith}):

$\Sigma_{V}(\epsilon,k)=\left(\frac{i}{2}\right)^{2}\frac{1}{(2\pi)^{4}}\int dk_{1}\int dk_{2}\int d\omega_{1}\int d\omega_{2}\, U_{q_{1}}U_{q_{2}}\times\left[\right.$\[
G_{0}^{R}(\epsilon-\omega_{2},k-k_{2})G_{0}^{K}(\epsilon-\omega_{1}-\omega_{2},k-k_{1}-k_{2})G_{0}^{K}(\epsilon-\omega_{1},k-k_{1})\]
\[
+G_{0}^{R}\epsilon-\omega_{2},k-k_{2})G_{0}^{A}(\epsilon-\omega_{1}-\omega_{2},k-k_{1}-k_{2})G_{0}^{R}(\epsilon-\omega_{1},k-k_{1})\]
\[
+G_{0}^{K}(\epsilon-\omega_{2},k-k_{2})G_{0}^{K}(\epsilon-\omega_{1}-\omega_{2},k-k_{1}-k_{2})G_{0}^{R}(\epsilon-\omega_{1},k-k_{1})\]
\begin{equation}
\left.+G_{0}^{K}(\epsilon-\omega_{2},k-k_{2})G_{0}^{A}(\epsilon-\omega_{1}-\omega_{2},k-k_{1}-k_{2})G_{0}^{K}(\epsilon-\omega_{1},k-k_{1})\right]\label{eq:}\end{equation}

\bibliographystyle{unsrt}
\bibliography{bib2,BibFM11}

\end{document}